\documentclass[12pt]{article}

\usepackage{scicite}
\usepackage{graphicx}
\usepackage{amsmath}
\usepackage{amssymb}
\usepackage{amsfonts}
\usepackage{tikz}
\usepackage{bm}% bold math
\usepackage{times}
\usepackage[version=4]{mhchem}
\usepackage[T1]{fontenc}
\usepackage[final]{pdfpages}

\newenvironment{sciabstract}{%
\begin{quote} \bf}
{\end{quote}}

\title{Frenkel biexcitons in hybrid HJ photophysical aggregates}
\author
{Elizabeth~Guti\'errez-Meza,$^{1}$ 
Ravyn~Malatesta,$^{1}$ 
Hongmo~Li,$^{2}$\\
Ilaria~Bargigia,$^{3}$
Ajay~Ram~Srimath~Kandada,$^{3}$\\  
David~A.~Valverde-Ch\'avez,$^{1}$ 
Seongmin~Kim,$^{2}$  
Hao~Li,$^{4}$\\
Natalie~Stingelin,$^{2,5}$ 
Sergei~Tretiak,$^{6}$\\ 
Eric~R.~Bittner,$^{4\dag}$ and 
Carlos~Silva-Acu\~na$^{1,2,7\ast}$\\
\\
\normalsize{$^{1}$School of Chemistry and Biochemistry, Georgia Institute of Technology,}\\
\normalsize{901 Atlantic Drive, Atlanta, GA~30332, United~States}\\
\normalsize{$^{2}$School of Materials Science and Engineering, Georgia Institute of Technology,}\\
\normalsize{North Avenue, Atlanta, GA~30332, United~States}\\
\normalsize{$^{3}$Department of Physics and Center for Functional Materials, Wake Forest University,}\\
\normalsize{1834 Wake Forest Road, Winston-Salem, NC~27109, United~States}\\
\normalsize{$^{4}$Department of Chemistry, University of Houston, Houston, Texas~77204, United~States}\\
\normalsize{$^{5}$ School of Chemical and Biomolecular Engineering,}\\
\normalsize{311 Ferst Drive NW, Atlanta, GA~30332, United~States}\\
\normalsize{$^{6}$Theoretical Division \& Center for Nonlinear Studies, Los Alamos National Laboratory,}\\
\normalsize{Los Alamos, NM~87545, United States}\\
\normalsize{$^{7}$School of Physics, Georgia Institute of Technology, 837 State Street,}\\
\normalsize{Atlanta, GA~30332, United~States}\\
\\
\normalsize{$^{\dag}$Corresponding author; E-mail address: ebittner@central.uh.edu}\\
\normalsize{$^{\ast}$Corresponding author; E-mail address:  carlos.silva@gatech.edu}
}

\date{}

\begin{document} 


\baselineskip24pt


\maketitle


\newpage
\begin{sciabstract}

Frenkel excitons, the primary photoexcitations in organic semiconductors that are unequivocally responsible for the optical properties of this materials class, are predicted to form \emph{bound} exciton pairs, {\em i.e.}, biexcitons. These are key intermediates, ubiquitous in many relevant photophysical processes; for example, they determine the exciton bimolecular annihilation dynamics in such systems. Deciphering the details of biexciton correlations is, thus, of utmost importance to understand the optical processes in these semiconductors. To date, however, due to their spectral ambiguity, there has been only scant direct evidence of bound biexcitons, limiting the insights that can be gained. Moreover, a quantum-mechanical basis describing biexciton correlation/stability has so far been lacking. By employing nonlinear coherent spectroscopy, we identify here bound biexcitons in a model polymeric semiconductor. We find, unexpectedly, that excitons with \emph{interchain} vibronic dispersion reveal \emph{intrachain} biexciton correlations and vice versa. Moreover, using a Frenkel exciton model, we can relate the biexciton binding energy to molecular parameters quantified by quantum chemistry, including the magnitude and sign of the exciton-exciton interaction the inter-site hopping energies. Therefore, our work promises a window towards general insights into the many-body electronic structure in polymeric semiconductors and beyond; e.g., other excitonic systems such as organic semiconductor crystals, molecular aggregates, photosynthetic light-harvesting complexes, or DNA.
\end{sciabstract}

\clearpage
\section*{Introduction}

The primary photoexcitations in organic semiconductors are molecular electronic singlet states (\ce{S_1}) termed Frenkel excitons~\cite{frenkel1931transformation}. Although these states are localized within a chromophore, at sufficiently high densities, exciton-exciton interactions start to dominate the optical properties of organic solids~\cite{agranovich1968collective}. %~\cite{agranovich2009excitations}. 
For example, the spontaneous formation of strongly-coupled light-matter quantum condensates in organic materials depends fundamentally on the details of exciton-exciton interactions~\cite{keeling2020bose}. Similarly, in biological light harvesting complexes, multi-exciton interactions may play important roles~\cite{scholes2011lessons}, while biexcitons can be crucial in cascade quantum emitters as a source of entangled photons~\cite{n2021mechanistic}. In order to fundamentally understand these processes, the molecular basis for biexciton stability needs to be established to provide a crucial comprehension of important structure/property interrelations that go beyond that achieved by examining single-quantum photoexcitations. Indeed, while excitons deliver an important window into electronic dispersion and its dependence on macromolecular structure and configuration, biexcitons provide a sophisticated probe of electronic structure  because they are consequential intermediates in a wide range of photophysical processes such as exciton dissociation into electrons (e$^{-}$) and holes (h$^+$)  
[\ce{S0 + 2 $\hbar\omega$ -> [2 S1]$^\ddagger$ -> 2e^- + 2h^+}]~\cite{silva2001efficient},
bimolecular annihilation 
[\ce{S1 + S1 -> [2 S1]$^\ddagger$ -> S0 + S0}]~\cite{stevens2001exciton}, 
and singlet fission producing triplet (\ce{T1}) states  
[\ce{S0 + 2 $\hbar\omega$ -> [2 S1]$^\ddagger$ -> T1 + T1}]~\cite{silva2002exciton}. 
Ref.~\citenum{stevens2001exciton} notes that bimolecular annihilation may be mediated both by resonance energy transfer and diffusion-limited exciton-exciton scattering, but in either case we invoke the key intermediate \ce{[2 S1]$^\ddagger$}.

While ample theoretical work points towards the existence of biexcitons in organic solids~\cite{spano1991biexciton,guo1995stable,gallagher1996theory,mazumdar1996exciton,agranovich2000kinematic,kun2009effect}  
and in optical lattices~\cite{xiang2012tunable}, there has been only indirect evidence of the dynamic formation of two-quantum exciton states in polymeric semiconductors by incoherent, sequential ultrafast excitation~\cite{silva2001efficient,stevens2001exciton,silva2002exciton}, in addition to a report on the \textit{indirect} observation of such species in a molecular aggregate~\cite{chakrabarti1998evidence} and a conjugated polymer~\cite{klimov1998biexcitons}. The reason is that the spectroscopic signatures of Frenkel biexcitons in organic semiconductors have thus far been ambiguous -- in strong contrast to Wannier-Mott biexcitons in quantum-confined semiconductor systems~\cite{stone2009exciton,turner2010coherent}. We note that Frenkel biexcitons have \textit{directly} been observed in solid argon~\cite{baba1991formation}. 

 Here, we report the \emph{direct} spectroscopic observation of \emph{bound} Frenkel biexcitons, {\em i.e.}, bound two-exciton quasiparticles (\ce{[2S1]$^\ddagger$}), in a  model polymeric semiconductor,  [poly(2,5-bis(3-hexadecylthiophene-2-yl)thieno[3,2-b]thiophene)]~\cite{mcculloch2006liquid} (PBTTT). This allows us to quantify biexciton binding energies. We selected PBTTT due to its thermotropic phase behavior, hosting liquid-crystalline phases at temperatures above ambient conditions~\cite{Chabinyc2007,Delongchamp2008}, which, we expect, renders the disordered energy landscape highly dynamic. Notably, we identify a spectral structure  that  is obscured within the inhomogeneously broadened linear excitation lineshape, which reflects both attractive and repulsive biexciton correlations and can be rationalized based on microscopic criteria. For this, we develop a Frenkel exciton model that quantifies how the balance of (i) exciton-exciton correlation energy, (ii) the energetic variations between different sites in the polymer aggregate, and (iii) exciton delocalization, define biexciton stability.   
%
%

%%%%%%%%%%%%%%%%%%%%%%%%%%%%%%%%%%%%%%%%%%%%%%%%%%%%%%%%%%%%%

\section*{Results}

In photophysical aggregates of semiconductor polymers, Frenkel excitons consist of 
chromophoric photoexcitations localized over at most a
few sub-units both 
along the polymer chain (categorized as a J-aggregate)
or between multiple polymer chains (H-aggregate)~\cite{Paquin:2013aa}, in many cases leading to hybrid HJ structures~\cite{Spano:2014aa}. 
Since the transition moment is typically oriented parallel to the polymer chains, and therefore interact in a head-to-tail fashion, J-aggregate excitons have a negative intrachain intersite hopping integral $t_{\mathrm{intra}}$, which is directly related to the free-exciton bandwidth $W=4|t_{\mathrm{intra}}|$ and is inversely related to the exciton effective mass $t_{\mathrm{intra}} = -\hbar^2/2\mu_{\mathrm{eff}}$. Therefore, $t_{\mathrm{intra}}$ determines exciton delocalization. In contrast to J-aggregates, H-aggregates feature a positive $t_{\mathrm{inter}}$. Both states (H- and J-aggregate states) are depicted in Fig.~\ref{fig:phased}, which shows the potential scenarios for {\em biexciton} interactions between pairs of excitons. (We note that $t_{\mathrm{intra}}$, $t_{\mathrm{inter}}$, and the exciton hopping term $t_{\mathrm{2Q,intra\,(inter)}}$ associated with exciton-exciton binding, are \emph{a priori} distinct; intra- and interhain exciton hopping terms certainly differ due to effects such as dipole-dipole orientation, charge-transfer contributions, and superexchange interactions, as do the corresponding hopping terms required for two-quantum binding. Within the effectively one-dimesnional model discussed below, we will denote the hopping integral simply as $t$.)  
%In this figure, t
The small ellipsoids represent the 
molecular sub-units, while the larger ellipsoids represent
single electron/hole excitons 
delocalized over multiple sites
either between chains (Fig.~\ref{fig:phased}A, B) or along the chains (Fig.~\ref{fig:phased}C, D).  
The latter of these form a quantum mechanical
basis for biexciton formation.

Exciton pairs become bound when their mutual attraction overcomes both the energy barrier imposed by local energetics  and their relative kinetic energy. However, there is a subtlety concerning the hopping integral $t$ and the energy strength of the exciton-exciton contact interaction $U$. These parameters \emph{must} have the {\em same} signs to produce a bound species. We note that \emph{both} attractive and repulsive interactions can produce bound states. 
%To see this, consider the solutions of the one-dimensional  Schr\"odinger equation with a $\delta$-function potential:
%\begin{align}
%    t \frac{\mathrm{d}^2 \psi(x)}{\mathrm{d} x^2}  + U\delta(x)\psi(x) = E\psi(x).
%    \label{eq:sch}
%\end{align}
%This system has a single bound state $\psi(x)$ which vanishes as $x\to\pm\infty$ and  has an exponential form $\psi(x)=\sqrt{\kappa}e^{-\kappa |x|}$, where $\kappa = U/2t$ is a positive constant and $E = t \kappa^2$. 
Ordinarily, we set $t<0$ and $U<0$ producing a single state that is energetically below %the $E>0$ scattering states.  
twice the exciton state. However, bound states can also occur when $t>0$ and $U>0$. Such states would occur at the above top of the energy band for free biexcitons. We further discuss this in the Supplemental Material. 

Curiously, H-aggregate excitons can move parallel to the 
chains with transfer integral $t<0$.
Moreover, the 
orientation of the local exciton dipoles give $U<0$,
implying that H-like exciton pairs can form bound biexcitons 
with an {\em attractive} binding energy {\em below} that of the 
energy for 
free H-like exciton pairs.  On the contrary, J-like excitons 
can move perpendicular to the chains with transfer integral
$t>0$ and with contact interaction $U>0$.  This implies that
the bound J-like biexciton states can form {\em above} the continuum for free
J-like exciton pairs. 
%Furthermore, a number of semiconducting polymers such as PBTTT that display 
Because of the \textit{hybrid} HJ aggregate nature~\cite{Spano:2014aa} established in PBTTT~\cite{hellmann2013controlling}, 
%in which 
both H-like and J-like behavior can be found offering the possibility of observing both 2J and 2H biexcitons within the same system. 

Additional insights can be gained considering that in a many-body framework, the exciton-exciton binding physics can be captured with a 1D Hubbard-like model with contact-interaction, $U$, between occupied nearest neighbors along the 1D chain. The hybrid HJ aggregate lattice is inherently two-dimensional, and it is possible that mixing between H-like and J-like excitons contributes to the formation of bound two-quantum quasiparticles. However, for the purposes of the analysis of this paper, we assume that the most important states for biexciton binding will involve the states with the longest and shortest wavevectors in each direction.  Based on this reasonable assumption, we can reduce the full 2D model to 1D by assuming separability of 
the single exciton motion between motion  along and perpendicular to the polymer chains and then introducing the exciton-exciton 
interactions. Introducing energetic disorder either into the local site energies or into the hopping integrals will break this separability and induce
mixing between between the H and J-like excitons. However, we assume that this mixing will be primarily within the bulk of the density of states and have modest impact (depending of course upon the relative magnitude of the disorder) on the extremal states corresponding to the bound biexcitons. Hence, we  argue that the 1D model captures the salient physics and provides the correct theoretical framework for interpreting and understanding the observations. We describe the Hamiltonian in the Supplemental Material along with its reduction from a more general 2D model implied by Fig.~\ref{fig:phased}. This model on the one hand accounts for energetic differences between repeating units, $\pm\Delta$, which we refer to as the {\em crenelation}. Although this effect gives rather artificial alternating site variations in energies, it is illustrative in that it captures the effects of disorder. 
For dissociated pairs, we can ignore the contact 
interaction and set $U=0$ to obtain the ratio of non-interacting excitons $E_{\mathrm{free}}$ and the intersite hopping integral $t$, $E_{\mathrm{free}}/t =2\sqrt{(\Delta/t)^2+4}$,
as the limiting case for two excitons with bulk momentum $k=0$.  
On the  other hand, the contact potential 
produces a single bound biexciton state with 
binding energy of  $E_b/t = (U/2t)^2$.
Exciton pairs are bound once 
$E_b < E_{\mathrm{free}}$ (for the 2J). This gives the insights that
the critical value of the coupling varies
as $\Delta^2$.
Hence, crenelation destabilizes the exciton binding suggesting that a highly crenelated lattice would disfavour biexciton binding.
Fig.~\ref{fig:system}E shows the 
two asymptotic limits of our model 
given as dashed curves, which 
results from a numerically exact diagonalization of 
a two-band Hubbard-like Hamiltonian.
For our 
numerical studies we set both $U<0$ and $t<0$ 
%corresponding to the XXX
so that the
bound 2H-biexciton are found
at the bottom of the energy spectrum. The 2J case
is identical except for a sign-change in the energy axis
producing a bound 2J state at the top of the energy spectrum 
(ee Supplemental Material for additional details). 

Another important point to take into account is that polymers are inherently heterogeneous systems. As a consequence, polymeric semiconductors  typically
exhibit 
a disordered energy landscape, 
which 
is reflected as a random variation in the
site energies in our model.  To examine this, we write
$\Delta_n = \bar\Delta+\delta\Delta_n$, where $
\overline{\delta\Delta_n\delta\Delta_{n'}} = \sigma_\Delta^2\delta_{nn'}$
gives the variance for a normal distribution with 
$\overline{\delta\Delta_n} = 0$.
Fig.~\ref{fig:phased}F shows the 
probability of biexciton formation ($P_{\mathrm{BE}}$) for various values of the
interaction strength $U$ and for increasing levels of site-wise energy fluctuations $\sigma_\Delta$ (for $\Delta/t=1$). The results indicate that bound biexcitons
are robust against local energy disorder
up to a certain point. Beyond this,
energetic disorder destabilizes the bound biexciton state. 
However, we also note that for extremely weakly interacting
excitons, local disorder can {\em enhance} the formation of
bound pairs when $U$ is close to the stability threshold. 
This can occur if %by chance 
neighboring sites have a fortuitously small 
enough energy off-set to form a local 
trapping site for a bound biexciton. 
However, the lineshape for these 
biexcitons will be broad to reflect the 
inhomogeneous distribution of their energies.

It is furthermore critical to emphasize that $U$ will depend on the local exciton dipole moment.  Fig.~\ref{fig:system}A shows front and side views of the chemical structure of
a PBTTT oligomer (three repeat units, a trimer) as optimized 
using Density Functional Theory (DFT, B3LYP/6-31G* level). 
The figure also displays the direction of the \ce{S_0 $\to$ S_1} transition dipole moment corresponding to the fundamental exciton state. 
The dipole, which is oriented along the long molecular axis, is large, corresponding to strong optical absorbance.  
Superimposed are the ground (\ce{S_0}) and excited (\ce{S_1}) state static dipoles, which have a small amplitude and are largely oriented perpendicular to the long molecular axis. These are highly sensitive to the instantaneous configurations adopted by the chain and average to zero over all chain conformations in the bulk system. 
Even though the conformational average vanishes, variance about this average is significant and maps into the local site energy variance $\sigma_\Delta$, thereby having an impact on the stability of multi-exciton states. 

Experimentally, we deduce from the linear spectral lineshape of PBTTT a dominant H-aggregate character (Fig.~\ref{fig:system}B). The reason is that the peak at $\sim 2.06$\,eV, which corresponds to the 0--0 vibronic peak, is suppressed with respect to the rest of the vibronic progression~\cite{Spano:2014aa}. 
This corresponds to the situation in Fig.~\ref{fig:phased}A
for a {\em single} exciton. 
 By fitting the absorption spectrum shown in Fig.~\ref{fig:system}B to a modified Frank-Condon progression~\cite{Clark:2009aa}, we obtained the energy of the main intramolecular vibration $E_p = 179$\,meV and %we can estimate 
 $W = 4|t_{\mathrm{inter}}|\approx 60$\,meV, expressed via the 0--0/0--1 absorbance ratio~\cite{Clark:2007aa}, and thus that $t_{\mathrm{inter}} \approx 15$\,meV.  We have previously demonstrated that predominantly J-aggregate character can be induced by blending the same material with polyethyleneoxide~\cite{hellmann2013controlling}. The 0--0 absorbance peak red-shifts by $\sim 70$--80\,meV in that case (see Supplemental Material for a comparison of the absorption lineshape of this batch of PBTTT processed as predominantly H- and J-aggregates).

In order to address the challenge of quantifying the \textit{biexciton} spectral structure, we performed coherent two-dimensional photoluminescence excitation (2D-PLE) spectroscopy on PBTTT.
Such measurements can directly identify biexciton resonances via two-quantum coherences (see details in Supplementary Material). They, thus, can quantify the biexciton binding energies~\cite{stone2009exciton,turner2010coherent}. 
This requires the construction of the coherent two-dimensional excitation spectrum via incoherent measurement of the time-integrated photoluminescence (PL) intensity due to a fourth-order excited-state population arising from the interference of wavepackets produced by a sequence of four light-matter interactions~\cite{Tekavec2007}, allowing the measurement of the spectral correlations between resonance involving pulses 1 and 2 and those corresponding to pulses 3 and 4 in Fig.~\ref{fig:Pulses_Feynman}A. Thereby, the two energy axes are constructed by Fourier transformation of the two-dimensional coherence decay function along time variables $t_{21}$ and $t_{43}$ at a fixed population time $t_{32}$. Accordingly, the spectral correlation along resulting energy axes ($\hbar \omega_{21}$, $\hbar \omega_{43}$) involve single-quantum ($|0\rangle \rightarrow |1\rangle$) and two-quantum  ($|0\rangle \rightrightarrows |2\rangle$) transitions, represented schematically in the left of Fig.~\ref{fig:Pulses_Feynman}B. 
The two principal coherent pathways involving the correlations between one- and two-quantum coherences under this detection scheme are depicted by the double-sided Feynman diagrams in Fig.~\ref{fig:Pulses_Feynman}B. We also display schematic diagrams of the one- and two-quantum 2D-PLE coherent spectra for two correlated transitions in  Figs.~\ref{fig:Pulses_Feynman}C and \ref{fig:Pulses_Feynman}D respectively. These may be correlated, for example, via a common ground state such as H-like and J-like states in a HJ aggregate, each evident in the diagonal of the 2D-PLE spectrum ($\hbar \omega_{21} =  \hbar \omega_{43}$). In addition, the spectral correlation between these two peaks is manifested by cross peaks between the diagonal features. Similarly, in the two-quantum correlation spectrum, a signal along the two-quantum diagonal ($\hbar \omega_{21-2Q} = 2 \hbar \omega_{43}$) signifies two non-interacting excitons, while a signal above or below the diagonal signifies binding interactions with repulsive or attractive character, depending on the signs of $t$ and $U$ (H- or J-like coupling), respectively. In our schematic in Fig.~\ref{fig:Pulses_Feynman}D, the lower energy diagonal peak displays 2J biexcitons (the two-quantum energy is higher than twice the one-quantum energy), while the higher-energy resonance displays structure 
corresponding to 2H biexcitons (in which the two-quantum energy is less than twice the one-quantum energy). 

We first focus on the single-quantum 2D-PLE lineshape in Fig.~\ref{fig:2Q_2D}A. 
We note that the spectral range  of the measurement, limited by the femtosecond laser spectrum displayed in Fig.~\ref{fig:system}B, overlaps with the entire 0--0 absorption peak. The 2D spectrum is dominated by a symmetric diagonal peak centered at $\hbar \omega_{21}=\hbar \omega_{43} \approx 2.06$\,eV, corresponding to the 0--0 peak energy.  
We also observe a weak diagonal signal centered at $\sim 1.99$\,eV, with even weaker structure identified at lower energy. These features display intense cross peaks with the (0--0) resonance. The rich spectral structure displayed at the low-energy tail of the 2D-PLE spectrum is not evident in the featureless linear absorption spectrum (Fig.~\ref{fig:system}) since it is obscured by the inhomogeneous lineshape, and demonstrates the existence of distinct states at the low-energy edge of the 0--0 absorption peak. We note that the energy of the weak diagonal feature corresponds to the 0--0 absorbance peak energy found in PBTTT J-aggregates induced when blending this semiconductor with a polar commodity plastic ~\cite{hellmann2013controlling}. 
We thus hypothesize that these are weak signatures of J-aggregate macromolecular conformations bearing effects of interactions with static dipoles (Fig.~\ref{fig:system}A).  We present in Supplemental Material the linear absorption spectrum of the same batch of PBTTT in a blend with an ionic liquid, processed such that the J-aggregate dominates the lineshape, which supports our assignment in Fig.~\ref{fig:2Q_2D}. The 0--0 absorption peak in that film is at 1.96\,eV, consistent with the spectrum reported in ref.~\citenum{hellmann2013controlling}, which supports the assignment of the weak, low-energy diagonal feature as the origin of the J-aggregate progression. 

We next examine two-quantum correlations within the spectral structure 
in Fig.~\ref{fig:2Q_2D}B, looking for exciton-exciton resonances that display repulsive or attractive biexciton binding energies as depicted in Fig.~\ref{fig:Pulses_Feynman}D. Most prominently, we observe a broad distribution along the two-quantum energy axis $\hbar \omega_{21,2Q}$ centered at the one-quantum energy $\hbar \omega_{43} = 2.06$\,eV (the 0--0 excitation maximum), with the peak of the distribution below the two-quantum diagonal, \textit{i.e.}, below the energy at which $ \hbar \omega_{21,2Q} = 2\hbar \omega_{43}$. Thereby, the spectrum cuts through the diagonal line, with a tail extending to the high-energy side of the diagonal. We also observe two-quantum peaks for the features at lower energy than the 0--0 peak in Fig.~\ref{fig:2Q_2D}A. These are both centered at higher energy than the diagonal; meaning that the two-quantum correlation is predominantly repulsive for the low-energy features.

%In order to extract biexciton binding energies from Fig.~\ref{fig:2Q_2D}C, i
In Fig.~\ref{fig:2Q_2D}D we display cuts of the two-quantum excitation spectrum at fixed one-quantum energies $\hbar \omega_{43}$. The cuts are  along $\hbar \omega_{21,2Q}$ relative to the two-quantum diagonal energy $\hbar \omega_{\mathrm{diag}}$. This is a good reference because signal on the diagonal corresponds to the energy of two excitons that coexist without interaction. %This then allows us to quantify biexciton binding energies. 
We discuss the structure in Fig.~\ref{fig:2Q_2D}D in detail in the Supplemental Material. Importantly, we notice, on the one hand, that the 0--0 absorption peak, assigned to the vibronic progression of an H aggregate~\cite{Spano:2014aa}, forms biexcitons with 
\textit{attractive} interactions with $E_{\mathrm{2H,b}} = -64 \pm 6$\,meV, corresponding to the situation in which $t_{\mathrm{2Q,intra}}<0, U<0$. On the other hand, the low-energy resonances, which we hypothesize to correspond to J-aggregate resonances, display predominantly repulsive two-quantum correlations, peaked at $E_{\mathrm{2J,b}} = +106 \pm 6$ and $+233 \pm 6$\,meV, respectively, which is produced when $t_{\mathrm{2Q,inter}}>0, U>0$. This supports the hypothesis presented in Fig.~\ref{fig:phased} that both attractive and repulsive biexciton binding coexist within the inhomogeneously broadened 0--0 linear absorption peak. 
%The two-quantum coherence spectrum in Fig.~\ref{fig:2Q_2D}C therefore reveals both 2H and 2J biexciton spectral structure. 
As importantly, using the experimentally determined $E_{\mathrm{2H,b}}$, corresponding to the correlated biexciton at the origin of the H-aggregate progression, and the estimate of $t_{\mathrm{inter}}$ from its linear lineshape in Fig.~\ref{fig:system}B, and under the hypothesis that $t_{\mathrm{inter}} \sim t_{\mathrm{2Q,intra}}$,  we determine that $U/t_{\mathrm{2Q,intra}} = 4.15$, placing the biexciton well into the range of stable biexcitons (see phase diagram in Supplemental Material). We find that while the scenario varies from sample to sample, this value is a lower limit, with an upper limit of $5.31$. We acknowledge that the hypothesis of the similarity of the interchain exciton hopping integral with that associated with exciton-exciton interactions is highly speculative, but it does permit us to begin to develop a framework to understand the physical basis for Frenkel biexciton binding. For the biexciton associated with the J-aggregate, we do not have a direct measurement of $t_{\mathrm{intra}}$, but we can speculate that it is a fraction of $t_{\mathrm{inter}}$ given the predominance of H-like character~\cite{Paquin:2013aa}, placing J-like biexcitons also well within the stable (but, surprisingly, repulsive) binding region, as depicted in Fig.~\ref{fig:phased}. Our analysis, thus, brings forth new opportunities to explore $U/t$ \textit{vs}.\ $\Delta/t$ phase diagrams for a broad class of molecular and macromolecular semiconductor materials, as portrayed in Fig.~\ref{fig:phased}E and in the Supplementary Material. For this, it is important to consider that $U/t$ can be manipulated by chemical design and processing %and, possibly, strong light coupling in cavities, 
while  $\Delta/t$ is expected to depend predominantly on the chemical structure of the polymer repeat unit. Some comparisons can already be made. Ref.~\citenum{chakrabarti1998evidence} gives, for instance, values of $t = 80$\,meV and $U/t \approx 4$ for epitaxially grown metal-halogen-phthalocyanine H-aggregates (fluoroaluminum phthalocyanine) based upon transient absorption measurements; no bound biexcitons were observed in a similar J-aggregate system (chloroindium phthalocyanine).

\section*{Discussion}

In summary, the complex spectral structure due to hybrid HJ excitons~\cite{Spano:2014aa} that are not resolved within the inhomogeneously-broadened linear absorption lineshape are revealed in a model polymeric semiconductor, PBTTT, using nonlinear coherent spectroscopy. All exciton resonances are identified along with their corresponding biexciton counterparts. A key finding is that that \emph{interchain} H-like excitons are associated with \emph{intrachain} exciton-exciton couplings. In contrast, \emph{intrachain} J-like excitons are paired by \emph{interchain} exciton-exciton couplings. In either case, the biexciton binding energy is related to the exciton-exciton contact interaction $U$ and the inter-site hopping energy $t$, which are intrinsic molecular parameters quantified by quantum chemistry and that are expected to be controllable via chemical design and polymer assembly. The reason is that $\Delta$ and $U$ depend on the local exciton dipole moment and, thus, likely on the  strength of the push-pull character of the polymer building blocks, \textit{i.e}. the chemistry of the monomer unit.  The insights presented in this report, thus, enable prediction of the stability of Frenkel biexcitons in polymeric semiconductors and provides insights  and design guidelines towards new materials discovery. More generally, it provides a platform to explore in unprecedented depth the many-body electronic structure of any material in which Frenkel excitons are the primary photoexcitations, ranging from organic semiconductors~\cite{agranovich2009excitations} and molecular crystals~\cite{hestand2017molecular} to photosynthetic light-harvesting complexes~\cite{scholes2011lessons} and protein/DNA structures~\cite{bittner2006lattice}.
   
\section*{Materials and Methods}

\subsection*{Sample preparation}

PBTTT [poly(2,5-bis(3-hexadecylthiophene-2-yl)thieno[3,2-b]thiophene)] was synthesized as previously reported ($M_n = 38$\,kg\,mol\textsuperscript{-1}, \DJ\textsubscript{M} = 1.78)~\cite{mcculloch2006liquid}. PBTTT was dissolved in 1,2-dichlorobenzene (o-DCB, Sigma-Aldrich) at $85^{\circ}$C and the solution was stirred for 1 hour. The concentration of the solution was 10\,mg\,mL\textsuperscript{-1}. Films were drop-cast at $30^{\circ}$C on sapphire substrates (UQG Optics, 10-mm diameter, 1-mm thickness) from as-prepared solution and were subsequently left to dry at this temperature.

In order to corroborate the assignment of the 0--0 J-aggregate absorption energy, we processed blends of PBTTT with the ionic liquid bis(trifluoromethylsulfonyl)imide (Iolitec), with an ionic liquid ratio of 2 per  repeating unit, in a solvent mixture consisting of o-DCB (Sigma-Aldrich) and cyclohexanone (CHN, Alfa Aesar) with a weight ratio of 4:6. The solutions were stirred at $110^{\circ}$C before casting. All chemicals described above were used without further purification. The films were cast from the solution on soda-lime glass slides at $80^{\circ}$C by wire-bar coating.

\subsection*{2D-PLE Spectroscopy}

The experimental setup shown schematically in Supplemental Material is derived from the phase-modulation and sensitive detection technique previously developed by  Marcus and coworkers~\cite{Tekavec2007}, and has been described in previous publications~\cite{vella2016ultrafast,Gregoire:2017aa}. It implements a sequence of four collinear femtosecond pulses that are generated in two Mach-Zehnder interferometers (MZI) that are nested in each arm of an outer MZI. Details are provided in Supplemental Material, including a detailed description of the experimental setup, a formal development of our approach for multi-quantum coherence measurements.

\subsection*{Density Functional Theory Calculations}

All DFT calculations were carried out with the Gaussian 16 (Revision A.03) software package~\cite{g16}, using the B3LYP hybrid functional and the 6–31G* basis set. The PBTTT chain was cut down into a trimer unit, and the hexadecyl side chains were replaced by ethyl groups in order to reduce the computational cost. The structure of oligomers were first optimized, and the ground-state dipole moment was obtained from single-point calculations. Excited states were calculated on top of the geometrical optimized structure using TD-DFT, and the excited state dipole moment and transition dipole moment were retrieved from the results corresponding to the first excited state before relaxation. The results are visualized using Avogadro~\cite{avgwebsite,hanwell2012avogadro}.

\section*{Supplementary Materials}

Supplementary material for this article is available at [url to be inserted].

%\bibliography{PBTTTlib}
%\bibliographystyle{Science}

\section*{Acknowledgments}
The authors are grateful to Prof.\ Martin Heeney for providing the material used for this work. \textbf{Funding: } %The work at Georgia Tech was funded by ... 
The work at Georgia Tech was funded by the National Science Foundation (DMR-1904293 [Silva] and DMREF-1729737 [Stingelin, Silva]). C.S.\ acknowledges support from the the School of Chemistry and Biochemistry and the College of Science at Georgia Tech. 
The work at the University of Houston was funded in
part by the  National Science Foundation (
%MRI-1531814,  % computer grant
CHE-1664971,   % single PI theory grant 
%CHE-1836080,  % EAGER/quantum light
DMR-1903785    % collaboration on perov.
) and the Robert A. Welch Foundation (E-1337). 
This work was also conducted in part at the Center for Integrated Nanotechnologies, a U.S.\ Department of Energy and Office of Basic Energy Science user facility. 
\textbf{Author contributions: } 
I.B.\ and A.R.S.K.\ constructed the 2D-PLE setup. 
The development of the demodulation technique to measure two-quantum coherent spectra was carried out by E.G.M., R.M., and D.A.V.C., aided by I.B.\ and supervised by C.S.A.\ and A.R.S.K.  
Hongmo L.\ and S.K. processed the samples, supervised by N.S. 
E.G.M\ measured and analyzed the 2D coherent excitation spectra, aided by D.A.V.C.\ and supervised by C.S.A. 
%D.A.V.C., and E.G.M.\ were responsible for the development of the the 2D-PIA setup, advised by A.R.S.K.\ and C.S.
Hongmo L.\ carried out the DFT calculations, supervised by S.T. 
Hao L. and E.R.B.\ developed the theoretical concepts and carried out the analysis. 
The intellectual basis for this paper was conceived by C.S.A., A.R.S.K., and E.R.B. 
All co-authors participated in the redaction of the manuscript. 
\textbf{Competing interests: } The authors declare no competing interests. 
\textbf{Data and materials availability: } All data needed to evaluate the conclusions in the paper are present in the paper and/or the Supplementary Materials.

%%%%%%%%%%%%%%%%%%%%% FIGURES %%%%%%%%%%%%%%%%%%%%%%%%%%%%%
\clearpage
\section*{Figures}

\begin{figure}[ht!]
   \centering
    \caption{
{\bf\sf Theoretical model for  biexciton formation.}
    ({\bf \sf A-D})
    H- vs J- biexciton configurations.  In A \& B, H-like excitons 
    are delocalized along the $\pi$-stacking direction  the polymer chains 
    and move along the chain with hopping integral
    $t<0$. Here, exciton dipoles are 
    aligned to produce an attractive $U<0$ contact interaction. 
    In comparison, the J-like excitons depicted in C \& D move
    between chains with $t>0$ and exciton dipole are aligned 
    co-facially to produce a repulsive $U>0$ contact interaction. 
    ({\bf \sf E}) Free vs. bound biexciton energies. 
    The energy of a free pair of excitons $E_{\mathrm{free}}$ is determined by their bandwidth, whereas the binding energy $E_B$ is determined by the contact interaction 
    $U$. Bound biexcitons become stable once $E_B<E_{\textrm{free}}$.
    Points correspond to energies from numerical 
    diagonalization while dashed curves
    are the asymptotic limits. The gray-dashed line is for two free 
    excitons on a homogeneous lattice ($\Delta=0$).
    ({\bf \sf F}) Probability for biexciton formation for 
    energetically disordered lattices. Here we set $\Delta/t = 1$
    and increase the variance in the local site energy. 
    Curves are labeled by the interaction $U$ for $U/t=2.5$ to $U/t=2.1$.}
    \label{fig:phased}
\end{figure}

\noindent

\begin{figure}[ht!]
    \caption{{\bf \sf Photophysical structure of PBTTT. }
    ({\bf \sf A}) Calculated orientation of the transition dipole moment (T), as well as the ground- and excited-state dipoles (GS, ES) for   %poly(2,5-bis(3-hexadecylthiophen-2-yl)thieno[3,2-\textit{b}]thiophene) (PBTTT)
    a trimer, for the particular configuration shown here. The angle of the ground-state dipole with respect to the transition dipole is 79.3$^{\circ}$ and that for the excited-state dipole is 99.5$^{\circ}$. The amplitudes of the ground state, excited state, and transition dipole moment are 0.98\,D, 0.93\,D, and 18.29\,D, respectively. The dipoles are not drawn to scale.
  ({\bf \sf B}) Linear absorption spectrum of PBTTT measured at 5\,K. Superimposed is the spectrum of the femtosecond pulse train used for the measurements presented in Fig.~\ref{fig:2Q_2D}. 
    }
    \label{fig:system}
\end{figure}

\begin{figure}[ht!]
    \centering
    \caption{{\bf\sf Two-dimensional coherent photoluminescence excitation (2D-PLE) spectroscopy.}  
    ({\bf \sf A}) Schematic of the experimental pulse sequence. 
    Here $\phi_{i}$ is the phase of pulse $i=1,2,3,4$. The inter-pulse delays are $t_{21}$ (coherence time), $t_{32}$ (population waiting time), and $t_{43}$ (coherence time). The phase-modulation reference waveforms for phase-sensitive detection are generated optically at frequencies $\Omega_{21}$ and $\Omega_{43}$, at which the relative phase $\phi_{21}=\phi_2-\phi_1$ and $\phi_{43}=\phi_4-\phi_3$ oscillate, respectively. In this work, time and spectrally integrated photoluminescence (PL) intensity is demodulated by phase-sensitive detection at the reference frequency $f_{\mathrm{ref}} = \Omega_{43} + \Omega_{21}$ and $ \Omega_{43} + 2\Omega_{21}$ for the one- and two-quantum correlation spectra, respectively, shown in Fig.~\ref{fig:2Q_2D}. We outline in Supplemental Material that the 2D-PLE lineshape can contain contributions from nonlinear incoherent population dynamics over the entire exciton lifetime, and that all spectral lineshapes presented in Fig.~\ref{fig:2Q_2D} are free from this undesired contribution under the excitation conditions of this experiment. 
    ({\bf \sf B}) Double-sided Feynman diagrams of the two most important two-quantum response terms that couple a ground state ($|0 \rangle$), a single exciton state ($|1 \rangle$), and a two-exciton state ($|2 \rangle$). 
    ({\bf \sf C}) Schematic representation of the 2D-PLE expected spectrum for two correlated optical transitions at energies $\hbar \omega_a$ and $\hbar \omega_b$. The spectral axes $\hbar \omega_{21}$ and $\hbar \omega_{43}$ are obtained by Fourier transform of the 2D coherent PL decay function along time variables $t_{21}$ and $t_{43}$ at fixed $t_{32}$. 
    ({\bf \sf D}) Schematic representation of a two-quantum 2D-PLE correlatios spectrum. The two-quantum energy $\hbar \omega_{21,2Q}$ corresponds to the two quantum coherences involving pulses 1 and 2, The diagonal line represents $\hbar \omega_{21,2Q} = 2 \hbar \omega_{43}$. }
    \label{fig:Pulses_Feynman}
\end{figure}

\begin{figure}[ht!]
    \centering
    \caption{{\bf \sf 2D-PLE spectra measured at 5\,K and $t_{32} = 30$\,fs.} ({\bf \sf A}) The real part of the one-quantum non-rephasing ($f_{\mathrm{ref}} = \Omega_{43} + \Omega_{21}$) spectrum. ({\bf \sf B}) The real part of the two-quantum ($f_{\mathrm{ref}} = \Omega_{43} + 2\Omega_{21}$)) spectrum.   ({\bf \sf C}) Diagrammatic representation involving the optical transitions associated with the interpretation of the spectrum in B. 
    ({\bf \sf D}) Spectral cuts along $ \hbar \omega_{21,2Q} - \hbar \omega_{\mathrm{diag}}$ spectral axis at fixed $\hbar \omega_{43} = 2.06$ (blue), $1.99$ (fuscia), and $1.94$\,eV (red) for the spectrum in B. Here $\hbar \omega_{\mathrm{diag}}$ is the $ \hbar \omega_{21,2Q} = 2\hbar \omega_{43}$ two-quantum diagonal energy, corresponding to zero net two-quantum correlation energy (neither binding nor repulsion). We chose to measure 2D coherent spectra with $t_{32}=30$\,fs in order to avoid ambiguous time ordering at $t_{32}=0$. Due to the highly transient nature of the two-quantum coherence signal, it would not be possible to measure with $t_{32}\gg 30$\,fs (see Supplemental Materials).}
    \label{fig:2Q_2D}
\end{figure}

%%%%%%%%%%%%%%%%%%%%%%%%% Actual figure Files %%%%%%%%%%%%%%%%%%%%%%%%%%%%%%%%%

\clearpage

{\bf \sf Figure 1}\\
\includegraphics[width=0.9\columnwidth]{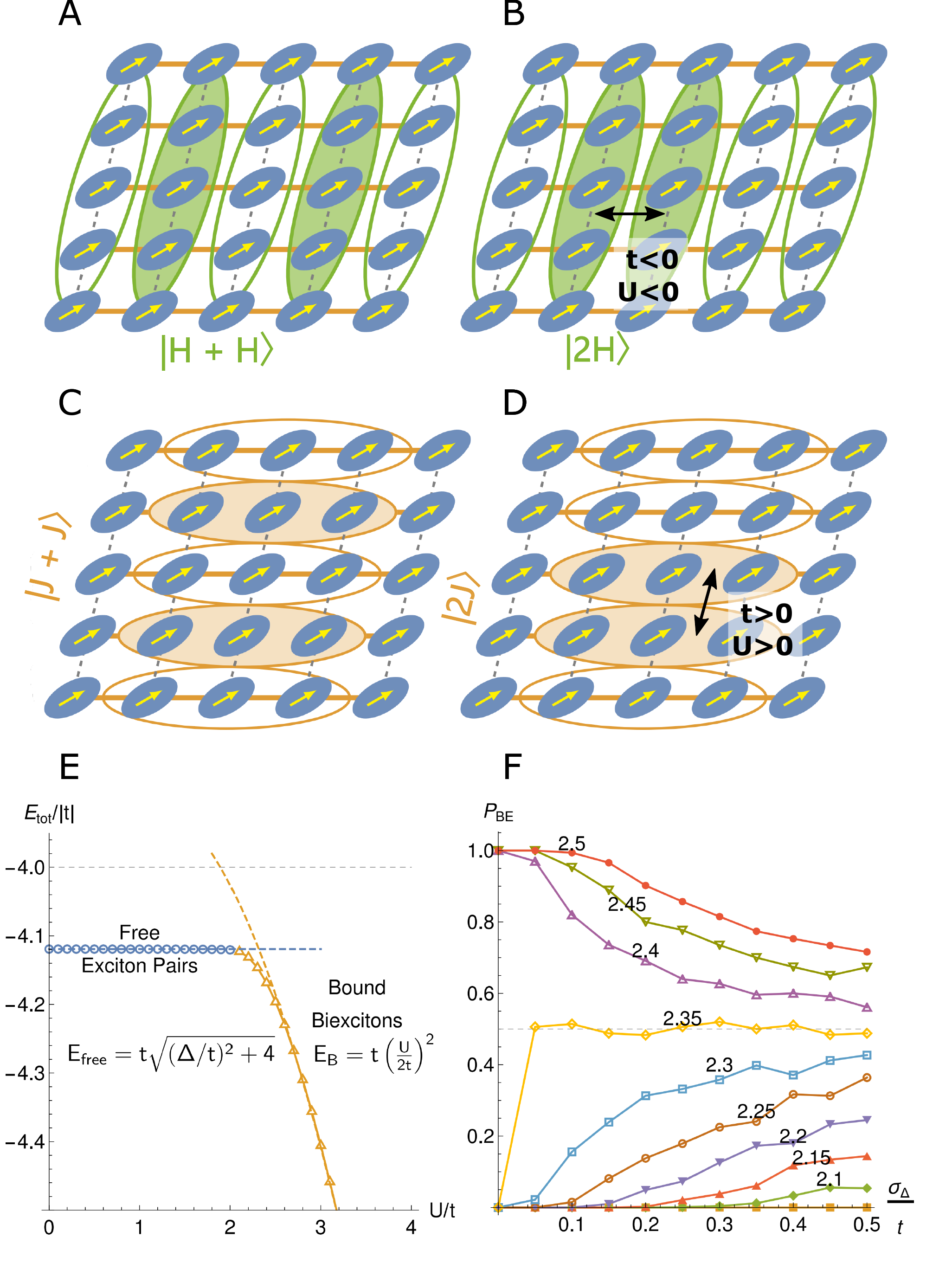}

\clearpage

{\bf \sf Figure 2}\\
\vspace{18pt}
\begin{tikzpicture}[thick,scale=0.75, every node/.style={scale=0.65}]

    \begin{scope}
     \node at (-3.5,3.5) {\bf \sf\Large A};
 %\node {\includegraphics[width=0.4\columnwidth]{Figures_Final/Figure1-structure.pdf}};
  \node {\includegraphics[width=0.9\columnwidth]{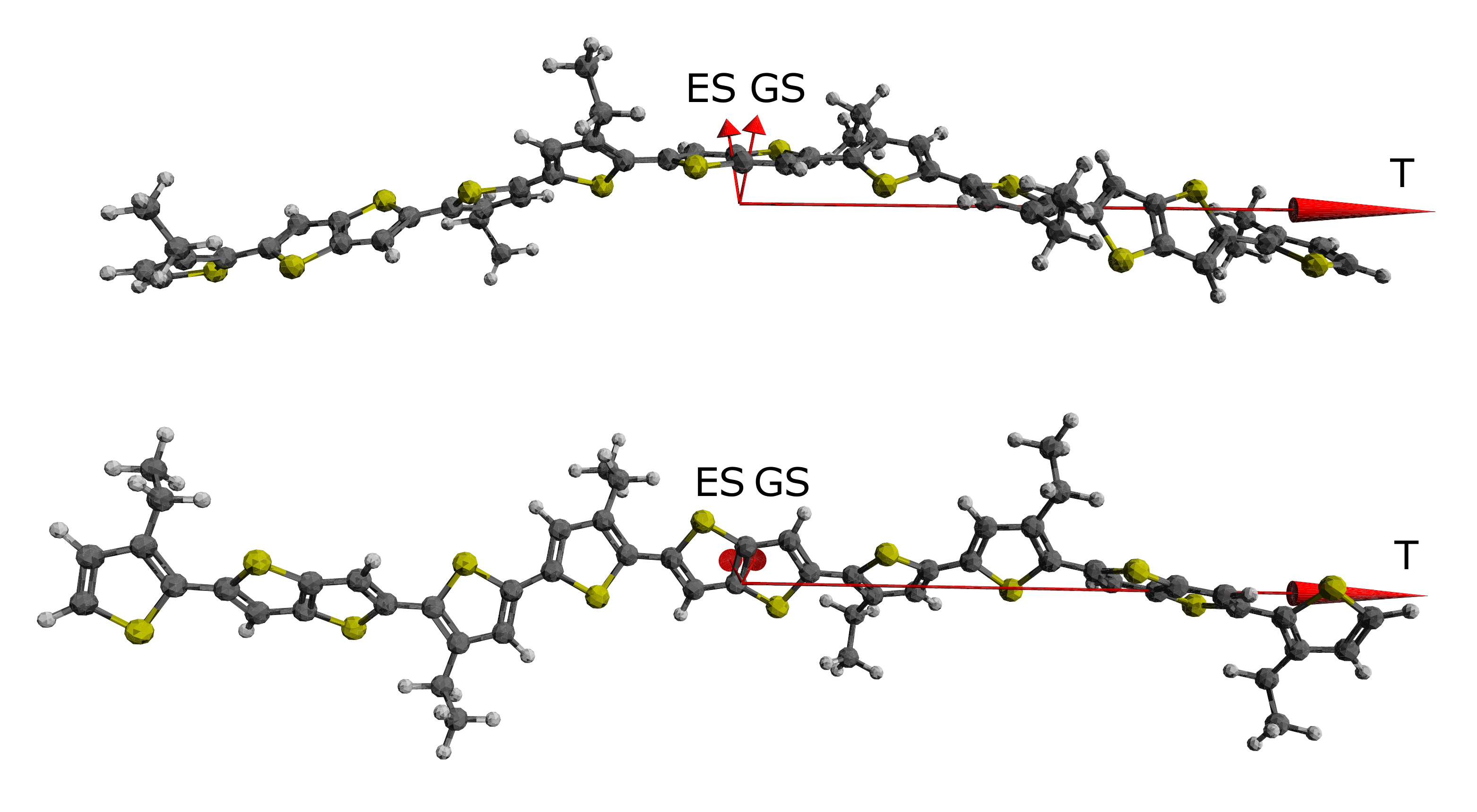}};
    \end{scope}

       \begin{scope}[shift={(12,0)}]
     \node at (-3.5,3.5) {\bf \sf\Large B};
 \node at (0,0) {\includegraphics[width=0.65\columnwidth]{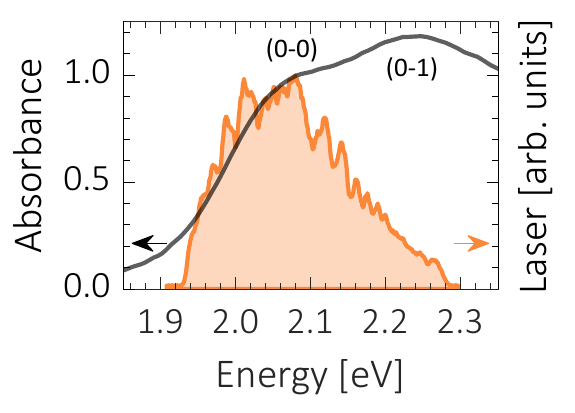}};
    \end{scope}
    \end{tikzpicture}
    
% %%%%%%%%%%%%%%%%%%%%%%%%%%%%%%%%%%%%%%%%%%

% %%%%%%%%%%%%%%%%%%%%%%%%%%%%%%%%%%%%%%%%%%%%%%%%%%%%%

\clearpage
{\bf \sf Figure 3}\\
    \includegraphics[width=\textwidth]{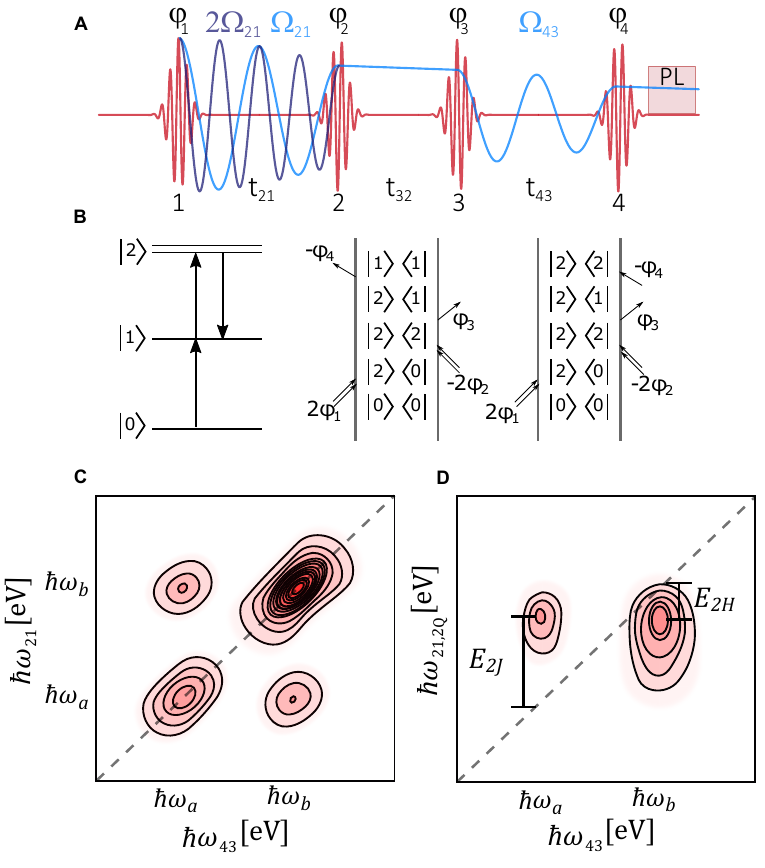}
   
%%%%%%%%%%%%%%%%%%%%%%%%%%%%%%%%%%%%%%%%%%%%%%%%%%%%%

\clearpage
{\bf \sf Figure 4}\\
\begin{center}
    \includegraphics[width=0.8\textwidth]{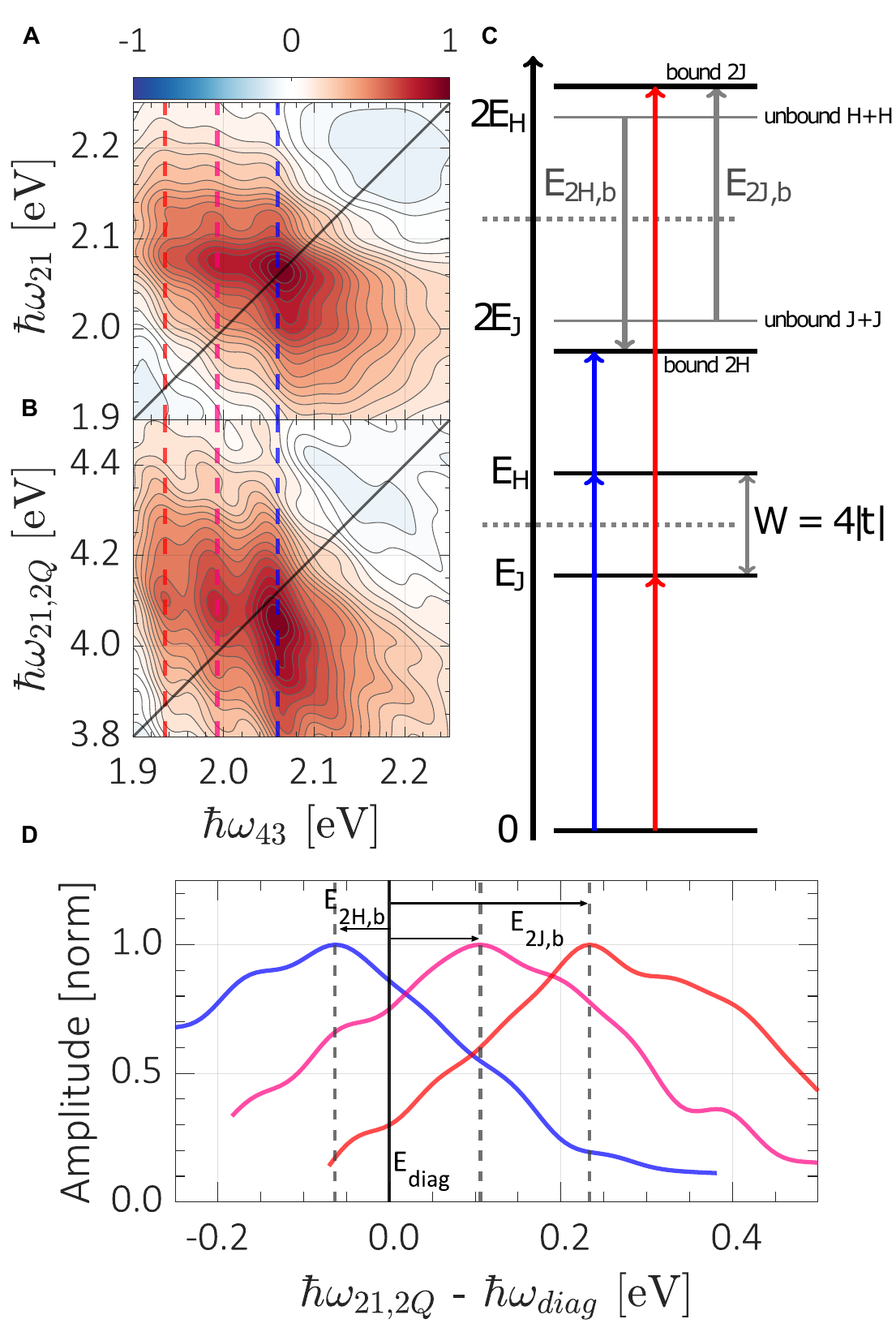}
\end{center}
%%%%%%%%%%%%%%%%%%%%%%%%%%%%%%%%%%%%%%%%%%%%%%%%%%%%%%%%

\newpage


\section*{Supplementary material}

# 1 2D-PLE Spectroscopy

## 1.1 Experimental Apparatus

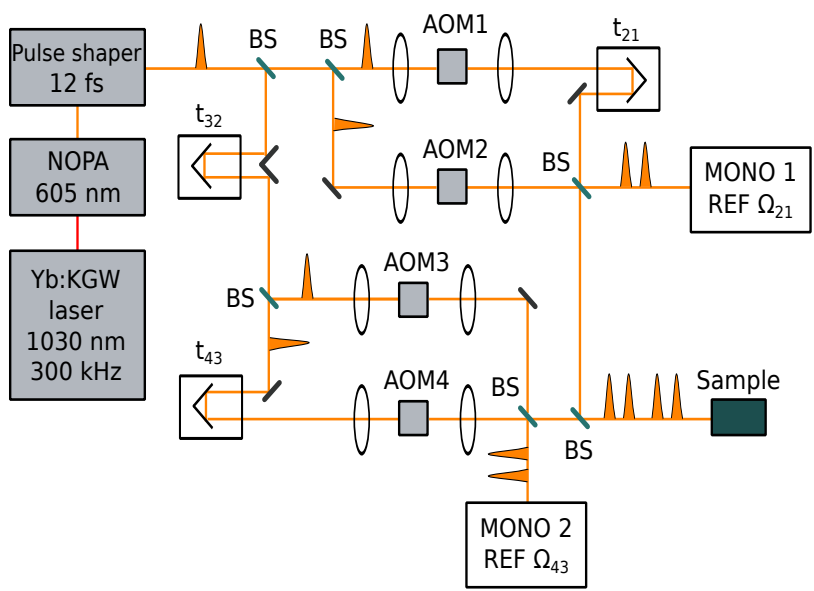

Figure S1: **Schematic diagram of the 2D-PLE set up.**
Yb:KGW: Ytterbium-doped potassium gadolinium tungstate laser. NOPA: Noncollinear optical parametric amplifier. BS: Beam splitter. AOM: Acousto-optic modulator. MONO: Monochromator. REF: Reference signal.

The experimental setup shown in Fig. S1 is derived from the phase-modulation and sensitive detection technique previously developed by Marcus and coworkers (*29*), and has been described in previous publications (*33,34*). It implements a sequence of four collinear femtosecond pulses that are generated in two Mach-Zehnder interferometers (MZI) that are nested into an outer MZI. We use a regenerative amplifier working at a repetition rate of 300 kHz, with a pulse train centered at 1030 nm and a pulse duration of 220 fs (Light Conversion Pharos Model

S2

25

PH1-20-0200-02-10). This output pumps a non-collinear optical parametric amplifier (NOPA) to produce 80-nm bandwidth pulses centered around 605 nm (Light Conversion Orpheus-N-3H). An adaptative $4f$ pulse shaper (BioPhotonics Solutions FemtoJock-P), pre-compresses the pulse to be 12 fs at the sample position by means of multiphoton intrapulse interference phase scan (MIIPS). The output is split by a 50% beam splitter. One arm is delayed with respect to the other using an optical delay line (Aerotech ANT130-060-L-25DU-MP) to define and control the inter-pulse delay, and then recombined in a collinear geometry with another identical beam splitter, defining the outer MZI. Each of these arms of this MZI is split into two sub-arms by 50% beam splitters, one of the inner arms in each MZI is delayed with respect to the other with an optical delay line (Aerotech ANT130-035-L-25DU-MP). and recombined to produce a four-pulse sequence. The phase-modulation scheme is achieved by placing an acousto-optic Bragg cell on each arm of the twin MZIs. The Bragg cells produce a frequency shift on each pulse that corresponds to its modulating frequency $\Omega_i$ ($i = 1, 2, 3, 4$), at around 200 MHz. The interfering excitation pulses generate a population signal oscillating at $\Omega_{21} = \Omega_2 - \Omega_1$ and $\Omega_{43} = \Omega_4 - \Omega_3$. In this work, we use $\Omega_{21} = 5\,\text{kHz}$ and $\Omega_{43} = 14\,\text{kHz}$. Within this phase modulation scheme, it is possible to isolate the fourth-order contribution to the signal by using a lock-in amplifier (Zurich Instruments UHFLI) with optically generated reference waveforms at $\Omega_{43} \pm \Omega_{21}$. These reference waveforms are optically constructed by means of avalanche photodiodes (APD) attached to the output of the two monochromators collecting half of the output of each inner MZI, as illustrated in Fig. S1.

The linear signals are obtained at the modulation frequencies $\Omega_{43}$ and $\Omega_{21}$. The rephasing and non-rephasing contributions to the fourth-order population are extracted by demodulating the signal at sideband frequencies $\Omega_{43} - \Omega_{21}$ and $\Omega_{43} + \Omega_{21}$, respectively, exploiting the multi-frequency and AM/FM modulation modules of the lockin amplifier. These four signals are acquired simultaneously.

S3

The fourth order population is detected with the time-integrated photoluminescence (PL) intensity, which is collected in a reflection geometry. The sample temperature is maintained at 5 K in a closed-cycle cryostation (Montana Instruments). The PL is isolated from the excitation laser by a long pass filter of 650 nm, detected by an APD (Hamamatsu C12703-01), and sent to the lock-in amplifier. The 2D maps are acquired by measuring the demodulated PL signal as a function of the time delays $t_{21}$ and $t_{43}$ (coherence times) at a fixed population time $t_{32}$. In this work the population time was set to $t_{32} = 30$ fs, in order to avoid any artifacts caused by the temporal overlap of the pulse pairs. The raw data obtained in the time-domain are then Fourier-transformed on both $t_{21}$ and $t_{43}$ variables to get 2D maps on the frequency domain at $\hbar\omega_{21}$ and $\hbar\omega_{43}$, leading to the rephasing and non-rephasing 2D maps.

### 1.2 Second-harmonic demodulation and two-quantum coherence

In order to isolate the two-quantum coherent signals, we based our experiments on the *n*th harmonic lock-in demodulation scheme developed by Bruder et al. (*38, 39*), using an inner MZI in Fig. S1. In this work, we demodulate at the second harmonic of the reference frequency $\Omega_{21}$ and combine it with the fundamental of $\Omega_{43}$, i.e., we demodulate the integrated PL intensity at the reference frequency $\Omega_{43} + 2\Omega_{21}$. As above, we Fourier-transform the measured two-dimensional coherence function along the variable $t_{21}$ and $t_{43}$ at a fixed value of $t_{32}$, and obtain a 2D spectrum with energy variables $\hbar\omega_{21,2Q}$ and $\hbar\omega_{43}$, where $\hbar\omega_{21,2Q}$ corresponds to the two-quantum energy axis. In what follows, we develop the formalism describing how this demodulation scheme results in the two-quantum correlation function of Fig. 4B in the main manuscript, following the initial development in ref. (*27*).

The total electric field $\vec{E}(t)$ is a sum of the four individual pulses $\vec{E}(t) = \vec{E}_1(t) + \vec{E}_2(t) + \vec{E}_3(t) + \vec{E}_4(t)$, where the field of the *j*th pulse ($j \in \{1, 2, 3, 4\}$) is defined as

$$\vec{E}_j(t) \equiv \vec{A}_j(t - t_j) \cos[\omega_L(t - t_j) + \phi_j], \tag{1}$$

S4

where $\vec{A}_j(t - t_j)$ is the temporal envelope of a pulse arriving at time $t_j$, $\omega_L$ is the laser carrier frequency, and $\phi_j$ is a constant phase. The Fourier transform yields the frequency-dependent field of pulse $j$

$$\hat{\vec{E}}_j(\omega) = \hat{\vec{E}}_j^+(\omega) + \hat{\vec{E}}_j^-(\omega), \tag{2}$$

where $\hat{\vec{E}}_j^+$ and $\hat{\vec{E}}_j^-$ are the forward and counter-rotating components of the Fourier transform respectively. The components have the explicit forms

$$\hat{\vec{E}}_j^+(\omega) = \vec{\alpha}_j(\omega)e^{i\omega t_j - i\phi_j}; \tag{3}$$

$$\hat{\vec{E}}_j^-(\omega) = \vec{\alpha}_j(-\omega)e^{i\omega t_j + i\phi_j}, \tag{4}$$

where $\vec{\alpha}_j(\omega)$ and $\phi_j$ are the spectral amplitude and phase of the $j$th pulse. From here on out we drop the hat, denoting each electric field with $\vec{E}_j$.

We consider the six pathways to an excited state population shown in Fig. S2 to contribute to the sum and difference frequency signals. In analogy to the Z-terms derived by Tekavec *et al.* (*27*) and using the sign convention of Grégoire (*40*), we derive six Z-terms using a perturbative expansion in the interaction picture. With $\hbar$ equal to 1, the first-order expansion of the population generated by interaction with pulse $j$ is

$$|\psi_j\rangle = i \sum_{r,s} \left[ \int_{-\infty}^{\infty} d\tau e^{i\omega_{rs}\tau} \vec{\mu}_{rs} \cdot \vec{E}_j(\tau)|r\rangle\langle s|g\rangle \right], \tag{5}$$

where $r, s \in \{|g\rangle, |1\rangle, |2\rangle\}$, where $|1\rangle$ and $|2\rangle$ denote one quantum and two quantum excited states respectively. We consider two possible two quantum excited states, the biexciton state $|b\rangle$ and the two-exciton state $|x\rangle$. Here we include the derivation of the term $Z_2 \propto \langle\psi_{431(1)}|\psi_{2(2)}\rangle$ and then provide the final Z-terms without their explicit derivations.

The term $Z_2$ corresponds to an excited population state in the second excited state generated by a fourth-order population in the bra and a second-order population in the ket. The bra is

S5

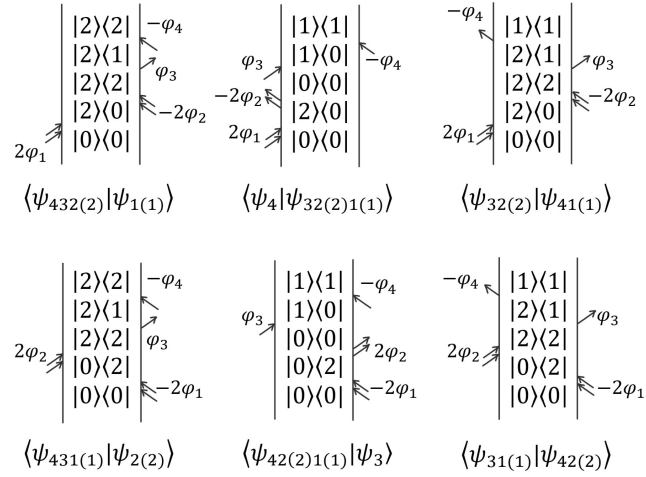

Figure S2: **Two-quantum excitation pathways.**
Double-sided Feynman diagrams for the pathways corresponding to (top) the three sum terms and (bottom) three difference terms.

excited in two cascaded interactions with the first pulse and subsequently de-excited then excited with pulses 3 and 4. The ket is excited in two cascaded interactions with the second pulse. There are four possible pathways matching this interaction order as shown in Fig. S3 where the final population state is either the biexciton, $|b\rangle$, or the two-exciton, $|x\rangle$, state. The second-order population generated by cascaded interactions with the second pulse is

$$\begin{aligned}
|\psi_{2(2)}\rangle &= i \sum_e \Big[ \int_{-\infty}^{\infty} d\tau \vec{\mu}_{e1} \cdot e^{i\omega_{e1}\tau} \vec{E}_2(\tau) \Big] \Big[ \int_{-\infty}^{\tau} d\tau' \vec{\mu}_{1g} \cdot e^{i\omega_{1g}\tau'} \vec{E}_2(\tau') \Big] |e\rangle \\
&\cong i \sum_e \Big[ \vec{\mu}_{e1} \cdot \vec{E}_2^-(\omega_{e1}) \Big] \Big[ \vec{\mu}_{1g} \cdot \vec{E}_2^-(\omega_{1g}) \Big] |e\rangle \\
&= i \sum_e [\vec{\mu}_{e1} \cdot \vec{\alpha}_2(-\omega_{e1})] [\vec{\mu}_{1g} \cdot \vec{\alpha}_2(-\omega_{1g})] e^{i(\omega_{e1}t_2+\phi_2)} e^{i(\omega_{1g}t_2+\phi_2)} |e\rangle \\
&= i \sum_e [\vec{\mu}_{e1} \cdot \vec{\alpha}_2(-\omega_{e1})] [\vec{\mu}_{1g} \cdot \vec{\alpha}_2(-\omega_{1g})] e^{i(\omega_{eg}t_2+2\phi_2)} |e\rangle,
\end{aligned} \qquad (6)$$

where the sum is over the two quantum excited states, $|b\rangle$ and $|x\rangle$.

S6

29

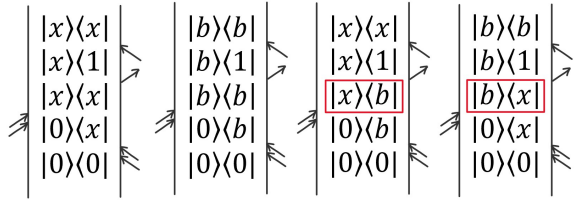

Figure S3: **Excitation pathway incorporating biexciton and two-exciton states.** Variations of the pathway of the $Z_2$ term incorporating the possible biexciton ($b$) and two-exciton ($x$) states. The last two diagrams show the dependence of $Z_2$ on the coupling between these states.

The fourth-order population generated by cascaded interactions with the first pulse followed by interactions with the third and fourth pulses is

$$|\psi_{431(1)}\rangle = -i \sum_{e,f} \Big[ \int_{-\infty}^{\infty} d\tau e^{i\omega_{e1}\tau} \vec{\mu}_{e1} \cdot \vec{E}_4(\tau) \Big] \Big[ \int_{-\infty}^{\tau} d\tau' e^{-i\omega_{f1}\tau'} \vec{\mu}_{1f} \cdot \vec{E}_3(\tau') \Big]$$
$$\times \Big[ \int_{-\infty}^{\tau'} d\tau'' \vec{\mu}_{f1} \cdot e^{i\omega_{f1}\tau''} \vec{E}_1(\tau'') \Big] \Big[ \int_{-\infty}^{\tau''} d\tau''' \vec{\mu}_{1g} \cdot e^{i\omega_{1g}\tau'''} \vec{E}_1(\tau''') \Big] |e\rangle, \quad (7)$$

where we enforce the time-ordering $t_1 < t_2 < t_3 < t_4$ and $e, f \in \{|b\rangle, |x\rangle\}$. With four independent Fourier transforms and the rotating wave approximation we find

$$|\psi_{431(1)}\rangle \cong -i \sum_{e,f} \Big[\vec{\mu}_{e1} \cdot \vec{E}_4^-(\omega_{e1})\Big] \Big[\vec{\mu}_{f1} \cdot \vec{E}_3^+(-\omega_{f1})\Big] \Big[\vec{\mu}_{f1} \cdot \vec{E}_1^-(\omega_{f1})\Big] \Big[\vec{\mu}_{1g} \cdot \vec{E}_1^-(\omega_{1g})\Big] |e\rangle$$
$$= -i \sum_{e,f} [\vec{\mu}_{e1} \cdot \vec{\alpha}_4(-\omega_{e1})] [\vec{\mu}_{f1} \cdot \vec{\alpha}_3(\omega_{f1})] [\vec{\mu}_{f1} \cdot \vec{\alpha}_1(-\omega_{f1})] [\vec{\mu}_{1g} \cdot \vec{\alpha}_1(-\omega_{1g})]$$
$$\times e^{i(\omega_{e1}t_4+\phi_4)} e^{-i(\omega_{f1}t_3+\phi_3)} e^{i(\omega_{fg}t_1+2\phi_1)} |e\rangle. \quad (8)$$

S7

Using $\vec{\alpha}(\omega) = \vec{\alpha}(-\omega)$ and $\vec{\alpha}_j \approx \vec{\alpha}_k$ for $j, k \in \{1, 2, 3, 4\}$, the inner product $\langle \psi_{431(1)} | \psi_{2(2)} \rangle$ gives

$$\begin{aligned}
Z_2 &= \langle \psi_{431(1)} | \psi_{2(2)} \rangle \\
&= e^{-i(\phi_{43} - 2\phi_{21})} \\
&\quad \times \Bigg[ |\vec{\mu}_{b1}|^4 |\vec{\mu}_{1g}|^2 |\vec{\alpha}(\omega_{b1})|^4 |\vec{\alpha}(\omega_{1g})|^2 e^{-i(\omega_{b1} t_{43} - \omega_{bg} t_{21})} \\
&\quad + |\vec{\mu}_{x1}|^4 |\vec{\mu}_{1g}|^2 |\vec{\alpha}(\omega_{x1})|^4 |\vec{\alpha}(\omega_{1g})|^2 e^{-i(\omega_{x1} t_{43} - \omega_{xg} t_{21})} \\
&\quad + |\vec{\mu}_{b1}|^2 |\vec{\mu}_{x1}|^2 |\vec{\mu}_{1g}|^2 |\vec{\alpha}(\omega_{b1})|^2 |\vec{\alpha}(\omega_{x1})|^2 |\vec{\alpha}(\omega_{1g})|^2 \\
&\quad \times \left[ e^{-i(\omega_{x1} t_{43} - \omega_{bg} t_{21} + \omega_{xb} t_{32})} + e^{-i(\omega_{b1} t_{43} - \omega_{xg} t_{21} + \omega_{bx} t_{32})} \right] \Bigg],
\end{aligned} \quad (9)$$

where the final two terms correspond to coupling pathways in which the system evolves during time $t_{32}$ as $|e\rangle\langle f|$ where $e \neq f$. This is the first of three difference terms in which the phase component is $e^{-i(\phi_{43} - 2\phi_{21})}$.

Acousto-optic modulation at the sum and difference frequencies $\phi_{43} + 2\phi_{21}$ and $\phi_{43} - 2\phi_{21}$ followed by demodulation using a lock-in amplifier allows for the simultaneous measurement of the sum and difference frequency signals for all interpulse delay times. From here on we denote the sum and difference frequencies as $2\Omega_{21} + \Omega_{43}$ and $-2\Omega_{21} + \Omega_{43}$, where $\Omega$ is used in place of $\phi$ to highlight the change to the experimental scheme. The phase modulation gives $\Omega_{21} = \Phi_{21} t' + \Omega_{21}^{\circ}$ and $\Omega_{43} = \Phi_{43} t' + \Omega_{43}^{\circ}$ where $t'$ is a quasicontinuous time variable and $\Omega_{21}^{\circ}$ and $\Omega_{43}^{\circ}$ are constant phases. The forward-rotating term for the frequency-dependent electric field of pulse $j$ with phase modulation becomes

$$\vec{E}_{m,j}^{+}(\omega) = \vec{\alpha}_j(\omega) e^{i\omega(t_j + mT) - i\Phi_j mT}, \quad (10)$$

where $m$ is the number of repetitions of pulse $j$ and $T$ is the time between each repetition and $mT = t'$. The phase component $\Omega_{m,j} = \Phi_j mT$ incrementally increases by $\Phi_j T$ with each successive repetition. Lock-in demodulation at the sum and difference frequencies are used to

S8

31

isolate the populations corresponding to each $Z$-term. The remaining $Z$-terms are

$$\begin{aligned}
Z_1 &= \langle \psi_{432(2)} | \psi_{1(1)} \rangle \\
&= e^{-i(\phi_{43}+2\phi_{21})} \\
&\quad \times \Bigg[ |\vec{\mu}_{b1}|^4 |\vec{\mu}_{1g}|^2 |\vec{\alpha}(\omega_{b1})|^4 |\vec{\alpha}(\omega_{1g})|^2 e^{-i(\omega_{b1} t_{43}+\omega_{bg} t_{21})} \\
&\quad + |\vec{\mu}_{x1}|^4 |\vec{\mu}_{1g}|^2 |\vec{\alpha}(\omega_{x1})|^4 |\vec{\alpha}(\omega_{1g})|^2 e^{-i(\omega_{x1} t_{43}+\omega_{xg} t_{21})} \\
&\quad + |\vec{\mu}_{b1}|^2 |\vec{\mu}_{x1}|^2 |\vec{\mu}_{1g}|^2 |\vec{\alpha}(\omega_{b1})|^2 |\vec{\alpha}(\omega_{x1})|^2 |\vec{\alpha}(\omega_{1g})|^2 \\
&\quad \times \Big[ e^{-i(\omega_{x1} t_{43}+\omega_{bg} t_{21}+\omega_{xb} t_{32})} + e^{-i(\omega_{b1} t_{43}+\omega_{xg} t_{21}+\omega_{bx} t_{32})} \Big] \Bigg],
\end{aligned} \quad (11)$$

$$\begin{aligned}
Z_3 &= \langle \psi_{42(2)1(1)} | \psi_3 \rangle \\
&= e^{-i(\phi_{43}-2\phi_{21})} \\
&\quad \times \Bigg[ |\vec{\mu}_{1g}|^4 |\vec{\mu}_{b1}|^2 |\vec{\alpha}(\omega_{1g})|^4 |\vec{\alpha}(\omega_{b1})|^2 e^{-i(\omega_{1g} t_{43}-\omega_{bg} t_{21})} \\
&\quad + |\vec{\mu}_{1g}|^4 |\vec{\mu}_{x1}|^2 |\vec{\alpha}(\omega_{1g})|^4 |\vec{\alpha}(\omega_{x1})|^2 e^{-i(\omega_{1g} t_{43}-\omega_{xg} t_{21})} \Bigg],
\end{aligned} \quad (12)$$

$$\begin{aligned}
Z_4 &= \langle \psi_4 | \psi_{32(2)1(1)} \rangle \\
&= e^{-i(\phi_{43}+2\phi_{21})} \\
&\quad \times \Bigg[ |\vec{\mu}_{1g}|^4 |\vec{\mu}_{b1}|^2 |\vec{\alpha}(\omega_{1g})|^4 |\vec{\alpha}(\omega_{b1})|^2 e^{-i(\omega_{1g} t_{43}+\omega_{bg} t_{21})} \\
&\quad + |\vec{\mu}_{1g}|^4 |\vec{\mu}_{x1}|^2 |\vec{\alpha}(\omega_{1g})|^4 |\vec{\alpha}(\omega_{x1})|^2 e^{-i(\omega_{1g} t_{43}+\omega_{xg} t_{21})} \Bigg],
\end{aligned} \quad (13)$$

$$Z_5 = \langle \psi_{32(2)} | \psi_{41(1)} \rangle = Z_1, \quad (14)$$

and

$$Z_6 = \langle \psi_{31(1)} | \psi_{42(2)} \rangle = Z_2, \quad (15)$$

S9

where $Z_5$ and $Z_6$ correspond to interactions of third-order populations.

The total sixth-order population can be divided between sum and difference terms

$$n^{(6)}(t_{43}, t_{32}, t_{21}) = n^{(6)}_{sum}(t_{43}, t_{32}, t_{21}) + n^{(6)}_{dif}(t_{43}, t_{32}, t_{21}), \tag{16}$$

where

$$n^{(6)}_{sum}(t_{43}, t_{32}, t_{21}) = \frac{2}{\hbar^6}\Re(2Z_1 + Z_4) \tag{17}$$

and

$$n^{(6)}_{dif}(t_{43}, t_{32}, t_{21}) = \frac{2}{\hbar^6}\Re(2Z_2 + Z_3). \tag{18}$$

Mukamel has criticized the scheme of Bruder et al. as being unselective for multi-quantum responses of non-interacting particles (*41*). He argued that for a particle ensemble, the photoluminescence signal reported in ref. 38 can be written as a simple product of one-particle coherences:

$$P^{(2)}(\tau) = N(N-1)|\mu_{eg}|^4 E_1^2 E_2^2 \exp(-2i(\omega_{eg} - \omega_1)\tau - 2i\phi_{21}), \tag{19}$$

with $\tau$ being the interpulse delay, and $\mu_{eg}$ the $|g\rangle \to |e\rangle$ transition dipole matrix element where $|e\rangle$ is the first excited state. Following the above equation where the signal detected at $2\phi_{21}$ is proportional to the excitation density and the electric field amplitudes, he argued that the measurement represents macroscopic coherence involving incoherent multi-particle interactions between spatially separated chromophores, which he considers to not carry any many-body information of the *microscopic* coherence. We consider that mechanism through which interactions between single-quantum excitations can be induced is synonymous with the incoherent population mixing (*42, 43*) that represents the non-linearity in the integrated PL signal, which we systematically rule out based on the arguments presented in Section 1.3. The second argument from Mukamel pertains specifically to the two-pulse excitation scenario. With two pulse

S10

33

interaction, one can reach distinct two-particle states - $|e_je_k\rangle$ and $|e_jg_k\rangle$, where $|e_{j/k}\rangle$ $|g_{j/k}\rangle$ represent the excited and ground states of a chromophore at jth/kth position within the ensemble of molecules. The PL however is then detected as the emission from the single particle state $|e_{j/k}\rangle$ whose population is not necessarily perturbed by the two-particle states and hence such a signal is not indicative of multi-quantum energy. Firstly, we can rule out the contribution from $|e_jg_k\rangle$ since we are demodulating the signal at the second harmonic. While in a two-pulse experiment, it may be true that the signatures of the biexciton population may not be detected at the exciton PL energy since there is not coherent pathway linking the two populations. It must be noted however that the PL is integrated over a long timescale allowing the biexciton population to relax to the single-quantum state with the coherent signatures still intact at the second-harmonic modulation frequency. Moreover, in the four-pulse experiment like the one in this work, the last two pulses create a coherent Louiville pathway to capture the biexciton coherence in the one-quantum population (for example, see the Feynman diagram for $\langle\psi_4|\psi_{32(2)1(1)}\rangle$ in Fig. S2). Based on these arguments and by the fact that we see features that represent biexciton binding and not simple exciton-pairs, we can conclude that we are indeed observing the *microscopic* multi-quantum coherences. Alternatively, spectroscopies based upon entangled photons may offer an unambiguous route for observing these coherences (*41, 44*).

### 1.3 Incoherent population mixing contributions

In condensed matter systems, photoexcitations are mobile and prone to inter-particle interactions such as exciton bimolecular annihilation, scattering processes and Auger recombination. These processes result in nonlinear population dynamics that arises from linear light-matter interactions and contributes as an incoherent signal to the measured lineshape (*42, 43*), masking the true 2D coherent response. According to Kalaee et al., these contributions can be differentiated from the true nonlinear signal by analyzing the phase and amplitude of the linear

S11

34

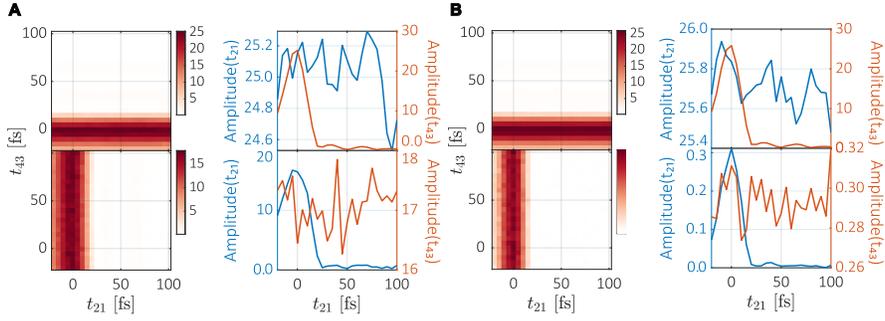

Figure S4: **Time-domain single-channel response measured with a fluence of $63\,\text{nJ}\,\text{cm}^{-2}$.**
Shown are the amplitude outputs of the lockin amplifier, obtained by scanning either $t_{21}$ or $t_{43}$, and stepping the other coherence time variable (left component of each panel). Also shown are their corresponding cuts along the $t_{21}$ direction (left axis in blue) and the $t_{43}$ direction (right axis in red). (**A**) Data obtained by demodulation at frequencies $\Omega_{43}$ (top panels) and $\Omega_{21}$ (bottom panels) and (**B**) $\Omega_{43}$ (top panels) and $2\Omega_{21}$ (bottom panels).

signal (*43*). The authors show that the incoherent-mixed signal have a $\pi$ phase shift with respect to the true nonlinear coherent signal. As pointed out in that work, the phase shift can be precisely identified if the linear and nonlinear signals are acquired simultaneously over the lock-in detection. This being the case in the measurements presented here, we performed this analysis and show the amplitude spectra of the linear and non-linear signals.

We measured the time-domain single-channel response, as shown in Fig. S4, by scanning along the $t_{21(43)}$ direction at fixed $t_{43(21)}$ to build the amplitude spectra displayed on the left component of panels (A) and (B), their corresponding cuts along $t_{21}$ and $t_{43}$ are also shown (see caption for details). These data were acquired with a pulse fluence of $63\,\text{nJ}\,\text{cm}^{-2}$, demodulating at frequencies $\Omega_{43}$ (top panels) and $\Omega_{21}$ (bottom panels) in (A) and at frequencies $\Omega_{43}$ (top panels) and $2\Omega_{21}$ (bottom panels) in (B). At frequency $\Omega_{43}$ (top panels) in (A) and (B), we observe a constant signal (in blue), when scanning along $t_{21}$ direction, as shown in the cut along

S12

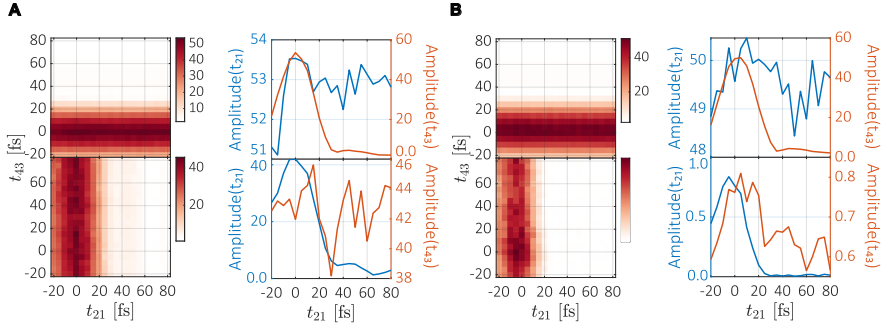

Figure S5: **Time-domain single-channel response measured with fluence of $251\,\mathrm{nJ\,cm^{-2}}$.** Shown are the amplitude outputs of the lockin amplifier, obtained by scanning either $t_{21}$ or $t_{43}$, and stepping the other coherence time variable (left component of each panel). Also shown are their corresponding cuts along the $t_{21}$ direction (left axis in blue) and the $t_{43}$ direction (right axis in red). (**A**) Data obtained by demodulation at frequencies $\Omega_{43}$ (top panels) and $\Omega_{21}$ (bottom panels) and (**B**) $\Omega_{43}$ (top panels) and $2\Omega_{21}$ (bottom panels).

this temporal axis, while a cut along $t_{43}$ direction shows a maximum amplitude intensity (in red). At frequencies $\Omega_{21}$ (bottom panels) in (A) and $2\Omega_{21}$ (bottom panels) in (B), we observe a constant signal (in red) when scanning along $t_{43}$ direction while a cut $t_{21}$ direction shows a maximum amplitude intensity (in blue). This behavior is expected in the absence of incoherent population mixing contributions to the signal, meaning that at the fluence used here, we got the true coherent response.

Conversely, the incoherent mixing contribution to the signal would be manifested by the presence of a cross-talk. To see this effect, let us turn our attention to Fig. S5, which shows the same measurements as in Fig. S4 but in this case acquired with a fluence of $251\,\mathrm{nJ\,cm^{-2}}$. In panel (A), which shows the demodulation at frequencies $\Omega_{43}$ (top panels) and $\Omega_{21}$ (bottom panels), the incoherent mixing contribution is not evident. However, in panel (B), the demodulation at frequency $2\Omega_{21}$ (bottom panels), shows a cross-talk in the amplitude spectrum and a

S13

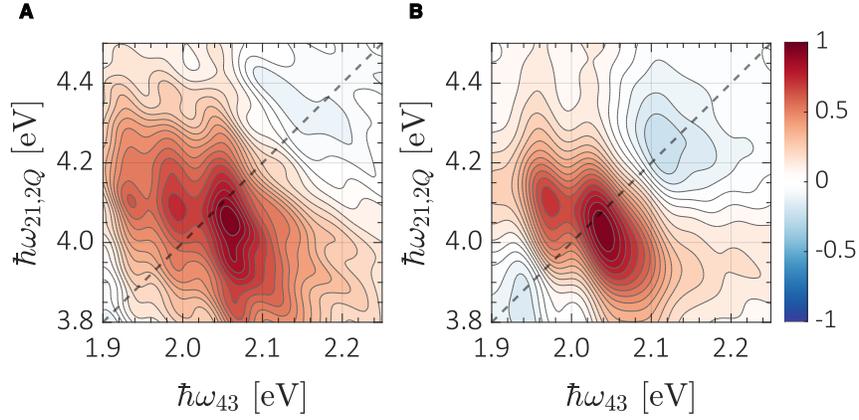

Figure S6: **Comparison of the real part of the 2Q 2D-PLE spectrum at different fluences.** Real part of the signal at reference frequency $f_{\text{ref}} = \Omega_{43} + 2\Omega_{21}$ at a fluence of (**A**) $63\,\text{nJ}\,\text{cm}^{-2}$ and (**B**) $251\,\text{nJ}\,\text{cm}^{-2}$, respectively.

maximum amplitude intensity along both $t_{21}$ and $t_{43}$ directions.

The results presented in Fig. S4 and Fig. S5 are significant not only because they allow us to identify the role of incoherent population mixing contributions at higher fluence, but they also to establish the optimal fluence range at which these contributions are not present. We also observed the effect of a high and low fluences on the 2D-PLE spectra lineshapes.

The comparison of the real part of the 2D-PLE spectrum, obtained by demodulating at frequency $\Omega_{43} + 2\Omega_{21}$ at fluences $63\,\text{nJ}\,\text{cm}^{-2}$ and $251\,\text{nJ}\,\text{cm}^{-2}$, is shown in Fig. S6. For a low fluence regime as shown in panel (A), we observe a well defined features that correspond to the two-quantum attractive and repulsive interactions as discussed in the main text. On the other hand, panel(B) shows the spectra acquired in the high fluence regime, which displays less structured features specially along the two-quantum energy axis $\hbar_{21,2Q}$. The pump spectra has a bit less bandwidth for the spectrum shown on panel (B), missing the lowest energy feature at

S14

1.94 eV; however, it is clear that incoherent mixing effects alter the lineshape on the 2D maps. It is important to underline that the results presented in this work were obtained in the low fluence regime, in order to avoid any incoherent mixing contribution.

## 2 Linear absorption spectra of H and J aggregates of PBTTT

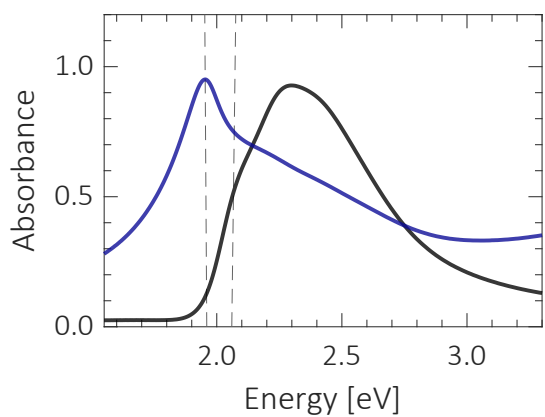

Figure S7: **Linear absorption spectra of neat PBTTT and a blend of PBTTT (black curve) with the ionic liquid bis(trifluoromethylsulfonyl)imid (blue curve).** The vertical dashed line corresponds to the 0–0 absorption peak of each spectrum, at 2.06 eV for the neat film and at 1.96 eV for the blend with the ionic liquid.

Fig. S7 displays linear absorption spectra of neat PBTTT processed similarly to the films presented in the main manuscript, and a blend with an ionic liquid. The peak energy of the 0–0 absorbance peak is at 2.06 eV as reported in Fig. 2, while that in the blend, which displays predominantly J-aggregate spectra structure, is at 1.96 eV. This supports the assignment of the diagonal structure in Fig. 4 to the 0–0 excitation resonance of H aggregates at 2.06 eV and of J aggregates at 1.99 eV.

S15

# 3 2D-PLE Supplementary Results

## 3.1 2D-PLE One- and Two-Quantum Spectra

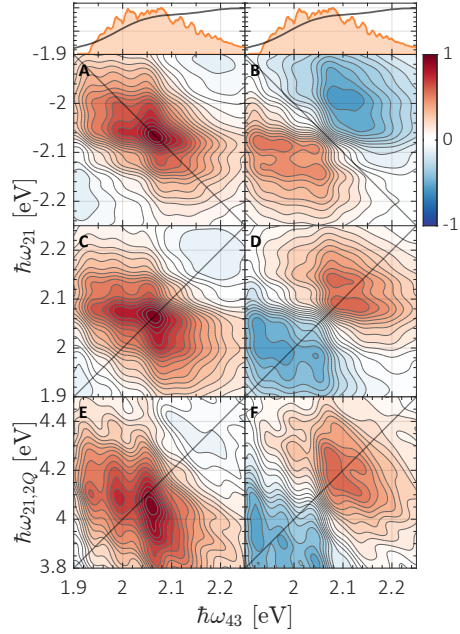

Figure S8: **One and two quantum 2D-PLE spectra measured at 5 K and $t_{32} = 30\,\text{fs}$.**
The laser spectrum (filled orange line) and the linear absorption spectrum (black line) are shown on the top panel for reference. Black line along the diagonal (**A, B**) The real and imaginary part of the one-quantum rephasing ($f_{\text{ref}} = \Omega_{43} - \Omega_{21}$) spectra. (**C, D**) The real and imaginary part of the one-quantum non-rephasing ($f_{\text{ref}} = \Omega_{43} + \Omega_{21}$) spectra. (**E, F**) The real and imaginary part of the two-quantum ($f_{\text{ref}} = \Omega_{43} + 2\Omega_{21}$)) spectra.

Fig. S8 shows the real and imaginary parts of the one-quantum rephasing( A, B), non-rephasing (C, D), and two-quantum coherence (E, F) phase combination of the 2D-PLE spectra at 5 K and population waiting time $t_{32} = 30\,\text{fs}$.

## 3.2 Spectral Analysis of Two-Quantum Spectrum

| $E$ (eV) | $E_{2H,b}$ (meV) | $E_{2J,b}$ (meV) |
|---|---|---|
| 2.06 | $-170 \pm 6$ | $138 \pm 8$ |
|  | $-64 \pm 6$ | $258 \pm 8$ |
| 1.99 | $-149 \pm 6$ | $106 \pm 6$ |
|  | $-57 \pm 6$ | $198 \pm 6$ |
|  |  | $276 \pm 6$ |
|  |  | $389 \pm 6$ |
| 1.94 | $-35 \pm 6$ | $53 \pm 8$ |
|  |  | $233 \pm 6$ |
|  |  | $329 \pm 8$ |
|  |  | $421 \pm 8$ |

Table 1: **Attractive and repulsive interaction energies** corresponding to Fig. 4 in the main text.

Table 1 displays the peak interaction energies associated with the spectral cuts in Fig. 4C in the main text.

## 3.3 Transient dynamics of two-quantum coherence signal

The spectrally integrated two-quantum population scan (red curve) shown in Fig. S9 was obtained by setting the coherence time delays $t_{21}$ and $t_{43}$ to zero and scanning along the time delay $t_{32}$. We compare this signal to the instrument response function corresponding to our 12-fs pulses.

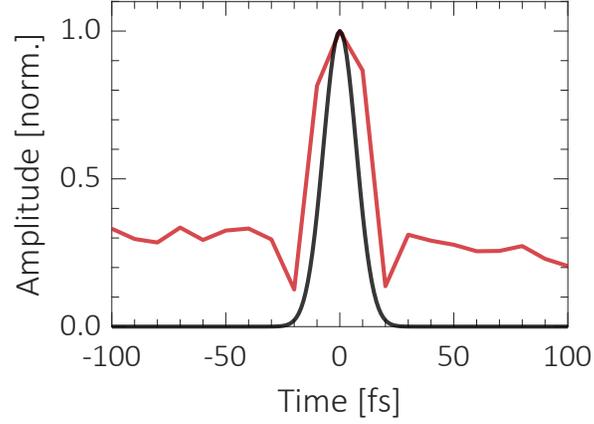

Figure S9: **Spectrally integrated population ($t_{32}$) scan of the two-quantum coherence signal (red curve) superimposed with the instrument response function (black curve). These data were measured with $t_{21} = t_{43} = 0$ fs.**

## 4 Supplementary Theoretical Modeling

### 4.1 Two-band Hubbard-like model

In a many-body framework, the exciton-exciton binding physics can be captured with a two-band Hubbard-like model with Hamiltonian

$$\hat{H} = E_0 + \Delta \sum_{j=1}^{N}(-1)^j \hat{c}_j^\dagger \hat{c}_j + t \sum_{j=1}^{N}(\hat{c}_j^\dagger \hat{c}_{j+1} + \hat{c}_{j+1}^\dagger \hat{c}_j) + U \sum_{j=1}^{N} \hat{c}_j^\dagger \hat{c}_j \hat{c}_{j+1}^\dagger \hat{c}_{j+1}. \quad (20)$$

The operators $\{\hat{c}_i, \hat{c}_j^\dagger\} = \delta_{ij}$ describe the creation and removal of an exciton in site $i$ and removal from site $j$, and in which odd-numbered sites correspond to sub-unit A and even-numbered sites correspond to sub-unit B. Periodic boundary conditions are such that $\hat{c}_N = \hat{c}_1$.

Formally, excitons are not Fermionic particles and are more precisely described as hard-core Bosons. However, the energy cost for placing two excitons on a given site is very high. Here this is enforced by Pauli exclusion which prevents any single site from being occupied

S18

by more than a single exciton. Our model accounts for energetic differences between repeating units, $\pm\Delta$, which we refer to as the *crenelation*. Parameter $t$ is the intersite hopping integral and $U$ is the exciton-exciton interaction energy, which we will take to be short ranged and due to contact interaction between excitons on neighboring sites. This model is consistent with the more elaborate one of Gallagher and Spano (*10*), but it captures the salient physics relevant to Frenkel biexciton stability and readily incorporates effects of crenelation, relevant to the incorporation fluctuations in site energy.

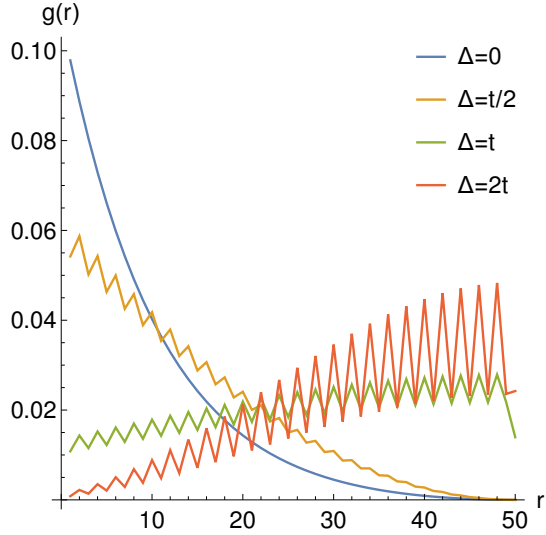

Figure S10: Bi-exciton pair distribution function for bound and un-bound pairs.

Numerical solutions were determined by directly diagonalizing $H$ for a given parameterization for systems up to 150 lattice sites. Fig. S10 shows the radial correlation function $g(r)$ between excitons and the total biexciton energy for various parametric choices of the tight-binding model discussed in the main paper. We define this such that value $g(0)$ is the probability of find-

S19

ing the two excitons on two directly adjacent sites since Pauli exclusion prevents two excitons for occupying the same site. For the homogeneous system ($\Delta = 0$), we see that biexcitons are stable only for $|U/t| > 2$. Below this, each exciton is delocalized over 1/2 the system.

For the homogeneous system, $\Delta = 0$, the exciton-exciton radial distribution falls off exponentially with increasing exciton/exciton separation, consistent with the continuum model given below. Recall that $\Delta$ is a measure of the energy difference between adjacent tight-binding sites and that increasing $\Delta$ introduces an energy cost for having two adjacent excitons, which we referred to as the crenelation. As $\Delta$ increases, the biexciton becomes unstable and dissociates into a pair of free excitons. Since our model calculations use a finite number of tight-binding sites (in this case 100) with periodic boundary conditions, the two "free" excitons are maximally separated to opposite sides of the periodic "ring". The oscillation in $g(r)$ reflects the fact that the odd-numbered sites have lower energy than the even-numbered sites and as such a single exciton is more likely to be in an odd-numbered sites.

Finally, we note that the exciton model we present is applicable to both $H$ and $J$-like coupling schemes. In the manuscript, we show results for $t < 0$ and $U < 0$, which are somewhat more intuitive to understand. When used for the $t > 0$ and $U > 0$ case, the bound biexciton will lie at the top of the energy spectrum above the continuum of scattering states rather than below the free-exciton continuum.

## 4.2 Justification for 1D reduced model

The model developed herein assumes separability between motion of the excitons parallel to the chain and perpendicular to the chains. In general, the physical system is not homogeneous/ordered lattice, but a more disordered aggregation of polymer chains. Nonetheless, we can use lattice models to gain general insight into the electronic states of these systems. To justify the 1D model, however, let us examine a generalized tight-binding model with Hamiltonian

S20

given by

$$H = \frac{1}{2}\sum_{nm} h_{nm}(\hat{c}_n^\dagger \hat{c}_m + \hat{c}_m^\dagger \hat{c}_n) + \sum_{nm} U_{nm}\hat{n}_m \hat{n}_n \qquad (21)$$

where $h_{nm}$ is the single particle term and $U_{nm}$ is the 2-body interaction. Operators $c_n$ and $c_n^\dagger$ remove and place excitations on sites indexed by $n$. These are Fermion operators with $\{c_n, c_m^\dagger\} = \delta_{nm}$ and we assume our excitons to be "spin-less" fermions such that Pauli exclusion precludes any site from being doubly excited. For particles with spin, the summations would also be over the spin-indices.

For the 2D model, we can assume a uniform square lattice of sites. For neighboring sites $n$ and $m$, $h_{nm} = t_J < 0$ if they are East/West neighbors and $h_{nm} = t_H > 0$ if they are North/South neighbors. Similarly, we assume the two-body term $U_{nm} = U_J < 0$ for East/West neighbors and $U_{nm} = U_H > 0$ for North/South neighbors, in accordance to the dipole alignments sketched in Fig. 1 of the paper. In principle, we can solve the many-particle Schrödinger equation numerically. For a 2-exciton system with $N$ total sites, the total number of 2-particle configurations is $N(N-1)/2$ and we estimate that one would need a $N = 50 \times 50 = 2500$ site lattice to compare with the 1D calculations presented in the paper. However, since we are ultimately interested in the two extremal states corresponding to the $|2J\rangle$ and $|2H\rangle$ states discussed in the main text, we make the following mathematical observations for the infinite-length system.

First, at $k_x = 0$, single particle states are fully delocalized along the E/W direction (corresponding to along the polymer chains) and have wavevector $k_y$ perpendicular to the chain (where $a$ is the lattice constant). Similarly, at $k_y = 0$, we have states totally delocalized in the N/S direction and with wavevector $k_x$ for their motions in the E/W direction. Let us write wavefunctions for two cases as

$$\phi_J(y) = \frac{1}{2\pi V} e^{ik_y y} \qquad (22)$$

$$\phi_H(x) = \frac{1}{2\pi V} e^{ik_x x}. \qquad (23)$$

S21

44

We see, then that the $\phi_J$ excitons have dispersion $E_J(k) = E_J + 2t_H \cos(k_y a)$, with effective mass $\mu_J < 0$ while the $\phi_H$ excitons have dispersion $E_H(k) = E_H - 2|t_J|\cos(k_x a)$ and effective mass $\mu_H > 0$.

Second, we can form 2-particle states using the $\phi_J$ and $\phi_H$ states. The form of the interaction implies that the two particles only interact at their contact. For the two cases delineated above, this corresponds to a line running E/W in the case of the JJ biexciton and line N/S for the HH biexciton. Integrating along these lines gives the biexciton interaction energy, which is simply $U_J$ and $U_H$. These two considerations allow one to deduce the 1D lattice model discussed in the text. Finally, one can move to a continuum model by writing the 1D lattice model in terms of the relative coordinate $r = a(n - m)$, and writing the contact interaction as $U(r) = U_0 \delta(r)$.

It is possible that mixed $H$ and $J$ biexcitons may be responsible for the small spectral features seen in the double-quantum coherence spectra. We leave this as a speculation to be explored in our future work.

### 4.3 Biexciton bound-states

Two excitons become bound when their mutual attraction is sufficient to overcome their kinetic energy. To examine this in the framework of the tight-binding model developed in manuscript, we take a continuum limit and work in a relative coordinate reference frame where $x = |r_1 - r_2|$ is the separation between two localized excitons. Thus, the Schrödinger equation can be approximated as

$$t\psi'' + U\delta(x)\psi = E\psi. \tag{24}$$

For bound states, $\psi(x)$ must vanish as $x \to \pm\infty$, giving that

$$\psi(x) = \begin{cases} \sqrt{\kappa}e^{-\kappa x} & x > 0 \\ \sqrt{\kappa}e^{+\kappa x} & x < 0 \end{cases} \tag{25}$$

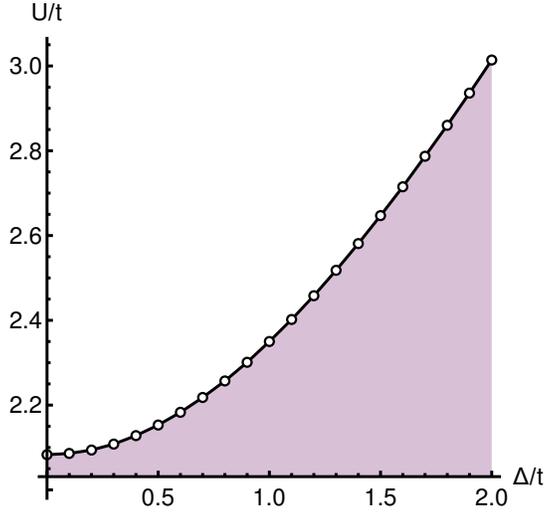

Figure S11: Phase diagram for biexciton formation.

where $\kappa = U/2t$ is a positive constant and $E = t\kappa^2$. In general, we take $t = -\hbar^2/2\mu_{eff}$ and $U < 0$ for an attractive potential giving rise to a bound state energetically *below* the continuum for the scattering states.

For molecular aggregates, the transfer integral $t$ depends upon the relative orientation of the transition moments between the exciton donor and acceptor sites and varies as $t = |\mu_{ge}|^2(1 - 3\cos^2(\theta))/R^3$ where $\theta$ is the relative orientation of the two transition moments and $R$ is the separation between the two species (taking $z$ to be along the direction connecting the two molecular centers). For $J$-aggregates, $\theta \approx 0$ for "head-to-tail" coaxial alignment of transition dipoles giving $t < 0$ and $\mu_{eff} > 0$ while for $H$-aggregates $\theta \approx \pi/2$ corresponding to the dipoles being aligned perpendicular to $z$ in which $t > 0$ and hence $\mu_{eff} < 0$. Optically, the bright state for a $J$-aggregate lies at the bottom of the energy band while the bright state for an $H$-aggregate lies at the top. This implies that two kinds of bound biexciton species could exist in systems such

S23

as PBTTT which is well-known for having both $H$ and $J$-like character.

### 4.4 Biexciton phase diagram

Fig. S11 shows a phase diagram for the formation of stable biexcitons based upon our model parameters. Above the parabolic curve, the system will form a bound (i.e. localized) biexciton state corresponding to a correlated exciton pair. The points correspond to results from the numerical solution of the two-exciton Schrödinger equation for a finite (20-site) lattice with periodic boundary conditions.



\end{document}